%% file: NMSSMplusRHN-scan-v2.tex
\documentclass[12pt]{article}

\usepackage[utf8]{inputenc}
\usepackage{cancel}
\usepackage{amssymb}
\usepackage{soul}
\usepackage{tabularx}
\usepackage{xcolor}
\usepackage{booktabs}
\usepackage{subcaption} 
\usepackage{colortbl}
\usepackage{authblk}
\DeclareUnicodeCharacter{00A0}{ }

\newcolumntype{C}{>{\centering\arraybackslash}X}
\input{defsV3}
\allowdisplaybreaks

\usepackage{array, makecell}
\usepackage{boldline}
\usepackage{cite}

\title{\bf{
Dark matter candidates in the NMSSM with RH neutrino superfields
}}
\author[a,b]{Daniel~E.~López-Fogliani\thanks{daniel.lopez@df.uba.ar}}
\author[c]{Andres~D.~Perez\thanks{andres.perez@iflp.unlp.edu.ar}}
\author[d]{Roberto~Ruiz~de~Austri\thanks{rruiz@ific.uv.es}}
\affil[a]{Instituto de Física de Buenos Aires UBA \& CONICET, Departamento de Física,
Facultad de Ciencia Exactas y Naturales, Universidad de Buenos Aires, 
1428 Buenos Aires, Argentina}
\affil[b]{
{Pontificia Universidad Católica Argentina, 
Av. Alicia Moreau de Justo 1500, 
1107~Buenos~Aires, Argentina}}
\affil[c]{IFLP, CONICET - Dpto. de Física, Universidad Nacional de La Plata,\protect\\
C.C. 67, 1900 La Plata, Argentina}
\affil[d]{Instituto de Física Corpuscular CSIC-UV, c/Catedrático José Beltrán 2, 46980 Paterna (Valencia), Spain}
\renewcommand\Affilfont{\itshape\small}

\date{}

\begin{document}
\maketitle
\begin{abstract}

R-parity conserving supersymmetric models with right-handed (RH) neutrinos are very appealing since they could naturally explain neutrino physics and also provide a good dark matter (DM) candidate such as the lightest supersymmetric particle (LSP). In this work we consider the next-to-minimal supersymmetric standard model (NMSSM) plus RH neutrino superfields, with effective Majorana masses dynamically generated at the electroweak scale (EW). We perform a scan of the relevant parameter space and study both possible DM candidates: RH sneutrino and neutralino. Especially for the case of RH sneutrino DM we analyse the intimate relation between both candidates to obtain the correct amount of relic density. 
Besides the well-known resonances, annihilations through scalar quartic couplings and coannihilation mechanisms with all kind of neutralinos, are crucial. Finally, we present the impact of current and future direct and indirect detection experiments on both DM candidates. 

\end{abstract}

{\small   Keywords: New Physics, Supersymmetry, Dark Matter, Sneutrino.} 

\newpage

\tableofcontents 


\section{Introduction}
\label{sec:intro}

Although the standard model (SM) is extraordinarily successful, there are still open questions in particle physics.
Cosmological observations reveal the presence of dark matter (DM) in the Universe and determine its abundance~\cite{Planck:2018VI,Bertone09,Feng:2010},
however the SM provides no viable candidate. Besides, the observation of neutrino oscillations phenomenon reveals that neutrinos must have extremely small, but non-vanishing masses~\cite{Pontecorvo:1969,FukudaSK:1998,Ahmed:2004,Araki:2005,Aliu:2005}, which are not allowed in the SM.

Introducing right-handed (RH) neutrinos is perhaps the simplest extension of the SM for describing neutrino masses and the observed neutrino oscillations\footnote{For a review about the origins of neutrino masses see Ref.~\cite{Mohapatra:2005wg} and references therein.}.
Although current results do not allow to establish the nature of the particle, the couplings of the RH neutrinos and the left-handed (LH) counterparts provide a source of Dirac- or Majorana-type mass term, depending on the model.
Small neutrino masses can be realized with very small Yukawa couplings, $O(10^{-13})$, in a Dirac-type way after electroweak symmetry breaking (EWSB)~\cite{Chang:1986bp,Mohapatra:1987hh,Fujikawa:1980yx,Cheng:1980qt}.
On the other hand, if we allow Majorana masses for the RH neutrinos, from the order of the  electroweak (EW) scale to the order of the Grand Unified Theory (GUT) scale, the smallness of the neutrino mass pattern can be achieve by a seesaw mechanism with neutrino Yukawa coupling from the order of the electron Yukawa coupling to order one ~\cite{Yanagida:1979as,Ramond:1979py,GellMann:1980vs,Yanagida:1980xy,Glashow1980}. 

In supersymmetric models (SUSY) the hierarchy problem is avoided, and the introduction of RH neutrino superfields can reproduce the neutrino physics. In addition, in models with $R$-parity conservation, the ones that are of interest for this work, besides the usual neutralino~\cite{Arcadi:2018,Roszkowski:2018}, an interesting alternative DM candidate arises as a by-product: the sneutrino, the scalar partner of the neutrino.

While LH sneutrino as DM is excluded by direct detection data, unless it makes up only a subdominant component of the DM~\cite{Falk:1994,Hall:1998,Arina:2007}, RH sneutrino as the lightest supersymmetric particle (LSP) can be a good DM candidate.
In the context of the Minimal Supersymmetric Standar Model (MSSM) viable thermal RH sneutrino DM requires mixing between the RH and LH sneutrinos with large SUSY breaking sneutrino trilinear parameters~\cite{Arina:2007,Arkani:2001,Belanger:2010,Dumont:2012,Kakizaki:2015,Banerjee:2016}, or/and lepton violating mass terms that split the sneutrino eigenstates~\cite{Hall:1998,Klapdor:2000,Kolb:2000,Asaka:2006,Asaka:2007}. Dirac sneutrinos as thermal DM could be achieved in models with an extended gauge group~\cite{Lee:2007mt,Belanger:2011rs,Romeri:2012}. On the other hand, models with small Yukawa couplings need non-thermal production of RH sneutrinos to account for the correct amount of DM~\cite{Gopalakrishna:2006,Page:2007,Choi:2018,Ghosh:2019}. In the recent work, Ref~\cite{Faber:2019}, different conditions and production mechanisms to obtain the measured amount of DM relic density have been studied. 

In this work we consider the next-to-minimal supersymmetric standard model (NMSSM)~\footnote{For an NMSSM review see for example Ref.~\cite{Ellwanger:2010,Maniatis:2009re}.} 
and extend it with three generations of RH neutrino superfields~\cite{Kitano:2000}.
In this context was first showed in~\cite{Cerdeno:2009} that thermal RH sneutrinos are good dark matter candidates.
The phenomenology of this model has been widely analysed including: the viability of the RH sneutrino as thermal DM with direct and indirect detection signatures~\cite{Cerdeno:2009,Deppisch:2008bp,Cerdeno2:2009,Demir:2010,Cerdeno:2011,Cerdeno2:2014,Cerdeno:2015,Cerdeno:2016,Chatterjee:2014,Ghosh2:2019,Cao:2019,Cao:2019qng}, the neutrino sector and DM candidates with spontaneous R-parity or CP violation~\cite{Kitano:2000,Huitu:2012,Huitu:2014,Tang:2015}, and collider and Higgs physics~\cite{Wang:2013,Cerdeno:2014,Cerdeno:2018,Cao:2019ofo}

The NMSSM includes in its formulation a singlet superfield to solve dynamically the $\mu$-problem of the MSSM. Moreover, including RH sneutrino superfields, extra terms in the superpotential between the RH neutrinos and the singlet superfield are allowed. This has a threefold effect in the neutrino and sneutrino sectors. First, the mass of the active neutrinos is generated by a seesaw mechanism, hence the neutrino Yukawa couplings can be of the same order as the electron Yukawa couplings. Second, the mass matrix of the sneutrino sector presents distinct features generated when the scalar singlet acquires vacuum expectation value (VEV) after EWSB, allowing a particular range for the mixing angle between LH and RH sneutrinos. Third, new decay and annihilation channels through the direct coupling to the singlet appear, contributing to the thermal production of sneutrino relic density.

In this paper, we present a low energy phenomenological SUSY realization with two DM candidates, 
the RH sneutrino and the usual neutralino. 
To explore the model, we perform a scan of the parameter space imposing constraints given by the measured amount of DM in the Universe, direct detection and indirect detection experiments of DM, 
SUSY and Higgs searches at colliders as well as neutrino physics.  
We identify the regions where either neutralinos or sneutrinos are the LSP and study the mechanisms to obtain the proper amount of relic density through thermal processes. As we will see, the sneutrino mainly uses three mechanisms: resonances with CP-even Higgs, annihilations through scalar quartic couplings with two CP-odd Higgs in the final state, and coannihilations with neutralinos. Thus, the relation between both DM candidates is explored in detail. Interestingly, coannihilations can be achieved with all kind of neutralinos, not only Higgsinos. 
Even more, the lightest slepton or colored particle, stau and stop respectively, can also take the role of the coannihilating partner.

We organize the paper as follows. In Section~\ref{sec:NMSSMplusRH}, we present the characteristics of the NMSSM plus RH neutrino superfields, and different mechanism to obtain the correct RH sneutrino relic density. Then, in Section~\ref{sec:scandetails} we explain the setup of the scan and the experimental constraints that have been considered. In Section~\ref{sec:scanresults}, we show the scan results, explore the main mechanisms to obtain the correct relic density, and discuss the characteristics of both DM candidates, with especial emphasis on the RH sneutrino. We also present several benchmark points for interesting cases. Finally, we show the impact of current direct and indirect detection constraints, and the important regions of the model that would be probed by next generation experiments. The conclusions are left for Section~\ref{sec:conclusions}.

\section{The NMSSM plus RH neutrino superfields}
\label{sec:NMSSMplusRH}

The next-to-minimal supersymmetric standard model (NMSSM) solves the $\mu$-problem of the minimal supersymmetric standard model (MSSM), but cannot explain the neutrino mass pattern. This can be solved including in its formulation RH neutrino superfields, generating dynamically Majorana masses at the EW scale. The superpotential considered is,
\bea
W &=& \epsilon_{\alpha \beta} \left( Y_e^{ij} \, \hat{H}_d^{\alpha} \, \hat{L}_i^{\beta} \, \hat{e}_j \, + \, Y_d^{ij} \, \hat{H}_d^{\alpha} \, \hat{Q}_i^{\beta} \, \hat{d}_j \, + \, Y_u^{ij} \, \hat{Q}_i^{\alpha} \, \hat{H}_u^{\beta} \, \hat{u}_j \, + \, Y_N^{ij} \, \hat{L}_i^{\alpha} \, \hat{H}_u^{\beta} \, \hat{N}_j \, + \, \lambda \, \hat{S} \, \hat{H}_u^{\alpha} \, \hat{H}_d^{\beta} \right) \nonumber \\ & + & \lambda_N^{ij} \, \hat{N}_i \, \hat{N}_j \, \hat{S} \, + \, \frac{\kappa}{3} \, \hat{S}^3,
\label{WNMSSM}
\eea
where $\hat{S}$ (L=0) is a singlet superfield, $\hat{N}$ (L=1) the neutrino superfield, $\epsilon_{\alpha \beta}$ ($\alpha, \beta = 1, 2$) is a totally antisymmetric tensor with $\epsilon_{12} =1$. As for the case of the NMSSM a $Z_3$ symmetry is invoked
to forbid the appearance of any dimensional parameter. The usual soft SUSY breaking terms in our case are
\bea
V_{soft}&=& \biggl[ \epsilon_{\alpha\beta}  \, \biggl( A_e^{ij} \, Y_e^{ij} \, H_d^{\alpha} \, \tilde{L}_i^{\beta} \, \tilde{e}_j \, + \, A_d^{ij} \, Y_d^{ij} \, H_d^{\alpha} \, \tilde{Q}_i^{\beta} \, \tilde{d}_j \, + \, A_u^{ij} \, Y_u^{ij} \, \tilde{Q}_i^{\alpha} \, H_u^{\beta} \, \tilde{u}_j \, + \, A_N^{ij} \, Y_N^{ij} \, \tilde{L}_i^{\alpha} \, H_u^{\beta} \, \tilde{N}_j \nonumber \\ & + & A_{\lambda} \, \lambda \, S \, H_u^{\alpha} \, H_d^{\beta} \biggr) \, + \, A_{\lambda_N}^{ij} \, \lambda_N^{ij} \, \tilde{N}_i \, \tilde{N}_j \, S \, + \, \frac{A_{\kappa} \, \kappa}{3} \, S^3 \biggr] \, + h.c. \nonumber \\ & + & m^2_{\phi_{ij}} \, \phi_i^{\dagger} \, \phi_j \, + \, m^2_{\theta_{ij}} \, \theta_i \, \theta_j^* \, + \, m^2_{H_d} \, H_d^{\dagger} \, H_d \, + \, m^2_{H_u} \, H_u^{\dagger} \, H_u \, + \, m^2_S \, S \, S^* \nonumber \\ & + & \frac{1}{2} \, M_1 \, \tilde{B} \, \tilde{B} \, + \, \frac{1}{2} \, M_2 \, \tilde{W}^i \, \tilde{W}^i \, + \, \frac{1}{2} \, M_3 \, \tilde{g}^a \, \tilde{g}^a,
\label{VNMSSM}
\eea
where $\phi ={\tilde{L},\tilde{Q}}$; $\theta ={\tilde{e},\tilde{N},\tilde{u},\tilde{d}}$ are the scalar components of the corresponding superfields, and the gauginos $\tilde{B}, \tilde{W}, \tilde{g}$, are the fermionic superpartners of the $B$, $W$ bosons, and gluons.

In this work we take all sfermion soft masses diagonal, $m^2_{ij}=m^2_{ii}=m^2_i$ and vanishing otherwise, were summation of repeated index convention was not used. Regarding the Yukawa and trilinear couplings, we assume that only the third generation of sfermions are non-zero, $T^{ij}=A^{ij}Y^{ij}$, without the summation convention, except in the neutrino case where $Y_{N}^{ij}$ and $A_N^{ij}$ are taken diagonal. Furthermore, we also consider diagonal the parameter $\lambda_N^{ij}=\lambda_N^{ii}=\lambda_N^{i}$, and its corresponding trilinear coupling, $A_{\lambda_N}^{ij} \, \lambda_N^{ij}=A_{\lambda_N}^{i} \, \lambda_N^{i} = T_{\lambda_N}^{i}$.

After electroweak symmetry breaking (EWSB) induced by the soft SUSY-breaking terms of $O(\text{TeV})$, and with the choice of CP conservation, the neutral Higgses ($H_{u,d}$) and the scalar singlet develop the following vacuum expectation values (VEVs)
\begin{equation}
\langle H_d \rangle = \frac{v_d}{\sqrt{2}}, \hspace{1cm} \langle H_u \rangle = \frac{v_u}{\sqrt{2}}, \hspace{1cm} \langle s \rangle = \frac{v_s}{\sqrt{2}},
\label{vevs}
\end{equation}
where $v^2=v_d^2+v_u^2= 4m_Z^2 /(g^2 + g'^2) \simeq (246 \text{ GeV})^2$, with $m_Z$ the $Z$ boson mass, and $g$ and $g'$ the $U(1)_Y$ and $SU(2)_L$ couplings, correspondingly. Then, the scalar components of the superfields $\hat{H_u}$, $\hat{H_d}$, and $\hat{S}$ can be written as
\bea
H_u=\left(\begin{array}{c}
 H_u^+\\
  \frac{v_u}{\sqrt{2}} + \frac{H_u^{\mathbb{R}} \, + \, i \, H_u^{\mathbb{I}}}{\sqrt{2}} 
  \end{array}\right),  \hspace{1cm}      H_d=\left(\begin{array}{c}
  \frac{v_d}{\sqrt{2}} + \frac{H_d^{\mathbb{R}} \, + \, i \, H_d^{\mathbb{I}}}{\sqrt{2}} \\
  H_d^-
  \end{array}\right),  \hspace{1cm}     S=\frac{v_s}{\sqrt{2}} + \frac{S^{\mathbb{R}} \, + \, i \, S^{\mathbb{I}}}{\sqrt{2}},
\eea
where the superscripts $\mathbb{R}$ and $\mathbb{I}$ indicate CP-even and CP-odd component fields, respectively.

In the basis ($H_d^{\mathbb{R}}$, $H_u^{\mathbb{R}}$, $S^{\mathbb{R}}$), we denote the CP-even scalar mass matrix as $M_S^2$. Likewise, dropping off the Goldstone mode, in the basis ($A$, $S^{\mathbb{I}}$), with $A=H_u^{\mathbb{I}} \cos \beta + H_d^{\mathbb{I}} \sin \beta$ and $\tan \beta = \frac{v_u}{v_d}$, the CP-odd scalar mass matrix is denoted $M_P^2$. The mass eigenstates of the CP-even Higgs $h_i$ with $i=1,2,3$, and the CP-odd Higgs $A_i$ with $i=1,2$ can be obtained by
\bea
\left(\begin{array}{c}
 h_1\\
 h_2\\
 h_3  
  \end{array}\right) = S_{ij} \left(\begin{array}{c}
 H_d^{\mathbb{R}}\\
 H_u^{\mathbb{R}}\\
 S^{\mathbb{R}} 
  \end{array}\right),      \hspace{2cm}     \left(\begin{array}{c}
 A_1\\
 A_2  
  \end{array}\right) = P_{ij} \left(\begin{array}{c}
 A\\
 S^{\mathbb{I}}
  \end{array}\right),
  \label{cphiggsmatrices}
\eea
where the matrices $S_{ij}$ and $P_{ij}$ diagonalize the mass matrices $M_S^2$ and $M_P^2$, respectively. The states are labeled according to the mass hierarchy $m_{h_1} < m_{h_2} < m_{h_3}$, and $m_{A_1} < m_{A_2}$. 
We can estimate $m_{A_i} \propto A_{\kappa} \, \kappa \, v_s$,
for the singlet dominated boson, which will be important later to allow RH sneutrino annihilations to light CP-odd scalars.

To generate an effective $\mu$-term, $\mu_{eff}=\frac{\lambda \, v_s}{\sqrt{2}}$, $v_s=O(\text{GeV-TeV})$ is needed. Then, the superpotential term $\lambda_N^{i} \, \hat{N}_i \, \hat{N}_i \, S$ generates dynamically a RH neutrino Majorana mass term $M_{N}^{i}=\frac{\lambda_N^{i} \, v_s}{\sqrt{2}} \sim O(\text{GeV-TeV})$, assuming that the parameters $\lambda$ and $\lambda_N^{i}$ are $O(0.1 - 1)$.

The LH neutrino masses, $m_{\nu_{L}}$, are generated by a seesaw mechanism. The general neutrino mass matrix is given by
\begin{equation}
M_{\nu} = 
\left(\begin{array}{cc}
 0 & m_D\\
  m_D^T & M_N
  \end{array}\right),
\end{equation}
therefore,
\bea
m_{\nu_{L}}\simeq - m_D \, M_N^{-1} \, m_D^T, 
\eea
with the dirac mass $m_D \simeq \frac{Y_N \, v_u}{\sqrt{2}}$. Thus $m_{\nu_{L}} \simeq \frac{Y_N^2 \, v_u^{2}}{\lambda_N \, v_s} \sim Y_N^2 \; \times$ EW scale, which implies $Y_N \sim 10^{-6}$ to get neutrino masses within the right order of magnitude.

The neutralino and chargino sectors are the same as in the NMSSM. The neutral colorless gauginos mix with the neutral higgsinos-singlinos and generate a symmetric 5$\times$5 mass matrix $M_{\chi^0}$. In the basis ($-i\tilde{B}$, $-i\tilde{W}^3$, $\tilde{H}_d^0$, $\tilde{H}_u^0$, $S$) we get
\bea
M_{\chi^0} = 
\left(\begin{array}{ccccc}
 M_1 & 0 & -\frac{g \, v_d}{2} & \frac{g \, v_u}{2} & 0\\
  0 & M_2 & \frac{g' \, v_d}{2} & -\frac{g' \, v_u}{2} & 0\\
  -\frac{g \, v_d}{2} & \frac{g' \, v_d}{2} & 0 & -\frac{\lambda \, v_s}{\sqrt{2}} & - \frac{\lambda \, v_u}{\sqrt{2}} \\
    \frac{g \, v_u}{2} & -\frac{g' \, v_u}{2} & -\frac{\lambda \, v_s}{\sqrt{2}} & 0 & - \frac{\lambda \, v_d}{\sqrt{2}} \\
    0 & 0 & - \frac{\lambda \, v_u}{\sqrt{2}} &- \frac{\lambda \, v_d}{\sqrt{2}} & \frac{2 \, \kappa \, v_s}{\sqrt{2}}
  \end{array}\right).
  \label{neutralinomatrix}
\eea

To obtain the mass eigenstates, the neutralino mass matrix can be diagonalized
\bea
N^* \, M_{\chi^0} \, N^{-1} = diag(m_{\chi_1^0},m_{\chi_2^0},m_{\chi_3^0},m_{\chi_4^0},m_{\chi_5^0}),
\label{diagneutralino}
\eea 
where $m_{\chi_1^0} < m_{\chi_2^0} < m_{\chi_3^0} < m_{\chi_4^0} < m_{\chi_5^0}$. Then, the neutralino eigenstates can be written as $\chi_i^0=N_{i1}\tilde{B} + N_{i2}\tilde{W}_3 + N_{i3}\tilde{H}_u^0 + N_{i4}\tilde{H}_d^0 + N_{i5}\tilde{S}$, with the matrix $N$ defining the composition of the neutralinos.

In the chargino sector, the charged Higgsinos and the charged gaugino mix forming two couples of physical chargino $\chi_1^{\pm}$ and $\chi_2^{\pm}$. In the basis ($\tilde{W}^{\pm}$, $\tilde{H}_{d,u}^{\pm}$) the chargino mass matrix is given by
\bea
M_{\chi^{\pm}} = 
\left(\begin{array}{cc}
 M_2 & \sqrt{2} \, \cos \beta \, m_W \\
  \sqrt{2} \, \sin \beta \, m_W & \mu 
  \end{array}\right).
  \label{charginomatrix}
\eea

Using two unitary matrices the chargino mass matrix can be diagonalized to obtain the mass eigenstates
 \bea
U^* \, M_{\chi^{\pm}} \, V^{-1} = diag(m_{\chi_1^{\pm}},m_{\chi_2^{\pm}}),
\label{diagchargino}
\eea 
where $m_{\chi_1^{\pm}} < m_{\chi_2^{\pm}}$.

\subsection{Sneutrino sector}
\label{sec:sneutrinosector}

The sneutrinos form a $12\times 12$ mass matrix divided into $6 \times 6$ submatrices
\begin{equation}
M_{\tilde{\nu}}^2 = 
\left(\begin{array}{cc}
 m_{\mathbb{R} \mathbb{R}}^2 & 0_{6\times 6}\\
  0_{6\times 6} & m_{\mathbb{I} \mathbb{I}}^2
  \end{array}\right),
\end{equation}
where again the subscripts $\mathbb{R}$ and $\mathbb{I}$ denote CP-even and CP-odd states, respectively. The off-diagonal submatrices are zero due to our choice of CP conservation. The submatrices are
\begin{equation}
m_{\mathbb{R} \mathbb{R}}^2 = 
\left(\begin{array}{cc}
 m_{L_{i}}^2 & A_{i}^{+}\\
  (A_{i}^{+})^T & m_{R_{i}}^2 + B_i
  \end{array}\right),  \hspace{1cm}  m_{\mathbb{I} \mathbb{I}}^2 = 
\left(\begin{array}{cc}
 m_{L_{i}}^2 & A_{i}^{-}\\
  (A_{i}^{-})^T & m_{R_{i}}^2 - B_i
  \end{array}\right),
\end{equation}
with
\bea
A_i^{+} &=& Y_N^{i} \, \left( \, A_N^i \, v_u \, + \, 2 \, \lambda_N^{i} \, v_u \, v_s \, - \, \lambda \, v_d \, v_s \, \right),\label{aplus}\\
A_i^{-} &=& Y_N^{i} \, \left( \, A_N^i \, v_u \, - \, 2 \, \lambda_N^{i} \, v_u \, v_s \, - \, \lambda \, v_d \, v_s \, \right),\label{aminus}\\
B_i &=& 2 \, \lambda_N^{i} \, \left( \, A_{\lambda_N}^{i} \, v_s \, + \, \kappa \, v_s^{2} \, - \, \lambda  \, v_u \, v_d \, \right),\label{bsneutrino}\\
m_{L_i}^2 &=& m_{\tilde{L}_i}^2 \, + \, (Y_N^{i})^2 \, v_u^2 \, + \, \frac{1}{2} \, m_Z^2 \, \cos 2 \beta ,\\
m_{R_i}^2 &=& m_{\tilde{N}_i}^2 \, + \, (Y_N^{i})^2 \, v_u^2 \, + \, 4 \, (\lambda_N^{i})^2 \, v_s^2,
\eea
as before, the index $i,j = 1,2,3$ are the family indices. The mixing between LH and RH sneutrinos is suppressed by the small neutrino Yukawa value in Eq.~(\ref{aplus}) and (\ref{aminus}). Considering only one family of neutrinos the mixing angle between LH and RH sneutrinos, $\theta_{\tilde{\nu}}$, can be approximated by
\bea
\tan 2 \theta_{\tilde{\nu}} & \simeq & \frac{2 \, A^{\pm}}{m_{L}^2 \, - \, \left( m_{R}^2 \, \pm \, B \right)} \\
& \simeq & \frac{2 \, Y_N \, \left( \, A_N \, v_u \, \pm \, 2 \, \lambda_N \, v_u \, v_s \, - \, \lambda \, v_d \, v_s \, \right)}{m_{\tilde{L}}^2 \, + \, \frac{1}{2} \, m_Z^2 \, \cos 2 \beta \, - \, m_{\tilde{N}}^2 \, - \, 4 \, \lambda_N^2 \, v_s^2 \, \mp \, 2 \, \lambda_N \, \left( \, A_{\lambda_N} \, v_s \, + \, \kappa \, v_s^{2} \, - \, \lambda  \, v_u \, v_d \, \right)}, \label{sneutmix1}
\eea
where the upper (lower) sign corresponds to the CP-even (CP-odd) state. For typical parameter values, $A_N \sim A_{\lambda_N} \sim O(\text{GeV})$, $\tan \beta \sim O(10)$, $\lambda \sim \kappa \sim \lambda_N$, we get
\bea
\tan 2 \theta_{\tilde{\nu}} & \sim & \frac{ Y_N \, \lambda_N \, v_s \, v_u}{m_{\tilde{L}}^2 \, + \, \frac{1}{2} \, m_Z^2 \, \cos 2 \beta \, - \, m_{\tilde{N}}^2 \, - \, \lambda_N^2 \, v_s^2 } \sim 10^{-2} \times Y_N \sim O(10^{-8}), \label{sneutmix2}
\eea
where we have used that $m_{\tilde{L}} \sim O(10^3)$ GeV to evade the stringent collider constraints on SUSY particles, and that $\lambda_N \, v_s \sim O(\text{EW})$ with $Y_N\sim 10^{-6}$ to reproduce the neutrino masses.

Due to the small mixing, the RH sneutrino masses can be taken as
\bea
m_{\tilde{\nu}_{R_i}}^2 \simeq m_{R_i}^2 \, \pm \, B_i,
\eea
here also the upper (lower) sign corresponds to the CP-even (CP-odd) state. We can see that the mass splitting is proportional to $\lambda_N^i$ in Eq.~(\ref{bsneutrino}). If $\lambda_N \rightarrow 0$ then $m_{\tilde{\nu}_{R_i}}^2 \simeq  m_{\tilde{N}_i}^2$. For typical parameter values in the NMSSM plus RH neutrinos
\bea
m_{\tilde{\nu}_{R_i}}^2 \approx m_{\tilde{N}_i}^2 + (2 \, \lambda_N^{i} \, v_s)^2 \pm \left( 2 \, T_{\lambda_N}^{i} \, v_s \, + \, 2 \, \lambda_N^{i} \, \kappa \, v_s^{2} \right).
\label{RHsneutrinomassAprox}
\eea
The new parameters with respect to the NMSSM $m_{\tilde{N}}^2$, $\lambda_N$ and $T_{\lambda_N}$ can be chosen to set the physical RH sneutrino mass, without affecting the rest of the mass spectrum. Moreover, the last two parameters also determine the RH sneutrino coupling to the singlet Higgs boson.

\subsection{RH sneutrino Dark Matter}
\label{sec:sneutrinoDM}

The existence of direct couplings of the RH sneutrino to Higgs bosons and neutralinos is a crucial feature of this model. The term $\lambda_N S N N$ in the superpotential, and the corresponding soft-breaking term $A_{\lambda_N}\lambda_N S N N$, 
generate the interactions through a mixing between the singlet and singlino components of $S$ with the CP-even Higgs bosons and neutralinos, respectively. 
Large values of $\lambda_N$ imply a more effective sneutrino annihilation, hence a smaller relic abundance. Moreover, the coupling also affects the value of the RH neutrino and sneutrino masses (see Eq.~(\ref{RHsneutrinomassAprox})), but not to the rest of the mass spectrum.


We denote $\tilde{\nu}_{R}$ a RH sneutrino,  $V$ a vector boson, $Z$ or $W$; $h_i$ ($A_i$) a neutral CP-even (CP-odd) scalar of the Higgs sector (recall that Higgs bosons are mixed with the scalar singlet); $f$ a SM fermion; $\nu_R$ a RH neutrino; $\chi_i^0$ a neutralino; $\chi_j^{\pm}$ a chargino; $\tilde{l}$ a slepton; $\tilde{q}$ a squark; and $\tilde{g}$ a gluino. The relevant annihilation channels involved to achieve the correct amount of RH sneutrino relic density are:

\begin{itemize}
\item $\tilde{\nu}_{R} \, \tilde{\nu}_{R} \rightarrow V V^*, V h_i, h_i h_i^{*}, A_i A_i^{*}, f \bar{f}, \nu_R \bar{\nu}_R$ via s-channel exchange of a CP-even Higgs boson.
\item $\tilde{\nu}_{R} \, \tilde{\nu}_{R} \rightarrow h_i h_i^{*}, A_i A_i^{*}$ via direct quartic coupling involving $\lambda$, $\kappa$ and $\lambda_N$.
\item $\tilde{\nu}_{R} \, \tilde{\nu}_{R} \rightarrow h_i h_i^{*}$ via t- and u-channel, exchanging a sneutrino.
\item $\tilde{\nu}_{R} \, \tilde{\nu}_{R} \rightarrow \nu_R \bar{\nu}_R$, via t- and u-channel exchanging a neutralino (via $\lambda_N \, \hat{N} \, \hat{N} \, \hat{S}$).
\item In the sneutrino sector, if CP-even and CP-odd states are nearly degenerate in mass, annihilation channels between them have to be taken into account, mainly via s-channel exchange of a CP-odd Higgs boson.
\item Coannihilation with neutralinos: $\tilde{\nu}_{R} \, \chi_i^0 \rightarrow$ lighter states, and $\chi_i^0 \, \chi_j^{\pm} \rightarrow$ lighter states. 
\item Coannihilation with sleptons, squarks, or gluinos: $\tilde{\nu}_{R} \, \tilde{l}_i \rightarrow$ lighter states, $\tilde{\nu}_{R} \, \tilde{q}_i \rightarrow$ lighter states, or $\tilde{\nu}_{R} \, \tilde{g} \rightarrow$ lighter states.
\end{itemize}


The viability of the annihilation mechanisms involving s-channel Higgs exchange as well as annihilation channels with scalar and pseudo-scalar Higgs bosons in the final state, depend on the mass hierarchy of the overall scalar sectors of the model. This relates the Higgs and sneutrino sectors. In the NMSSM both CP-even and CP-odd Higgs states can be very light with significant singlet component, making these annihilation channels kinematically allowed for very light sneutrino masses. Thus, an allowed DM relic density can be obtained for low mass DM candidates that would otherwise be excluded. As usual, the annihilation processes that involve s-channel Higgs exchange are enhanced near the resonant mass condition, $m_{\tilde{\nu}_R} \simeq m_{h_i}/2$.  We will see that resonances and direct annihilations through quartic coupling to a pair of pseudo-scalar Higgs are especially relevant for low mass sneutrinos.

On the other hand, away from resonances the direct annihilation of RH sneutrinos is not efficient enough leading to an overproduction of DM.
However if the mass splitting between the RH sneutrinos and other SUSY particles is small coannihilations can be efficient to keep the thermal equilibrium for longer and therefore their relic abundance can fulfill observations \cite{Griest:1990kh}.

As RH sneutrino couplings with neutralinos is mainly through a Higgs boson exchange (for example using the terms $\lambda_N  N  N  S$ and $\lambda  S  H  H$), coannihilations with neutralinos mostly Higgsino are more efficient. 
In this case, the allowed sneutrino mass range will inherit the constraints on the neutralino (and chargino) sectors regarding its masses, 
i.e. we get $m_{\tilde{\nu}_N} \gsim 100$ GeV due to collider searches on SUSY particles. As we will see in detail later, this is one of the main mechanisms to obtain a correct relic density. However, coannihilations with all kind of neutralinos are possible, allowing to find viable RH sneutrino DM for a wide range of parameters.

\subsection{RH sneutrino direct detection}
\label{sec:RHsneutrinoDD}

For RH sneutrino, the scattering with a nucleon $N$, either $N=p,n$, at tree level only occurs via $t$-channel exchange of a neutral CP-even Higgs boson. In the non-relativistic regime, the effective operator $L_{\widetilde{\nu}_R-N}^{eff} \; = \; g_N \widetilde{\nu}_R \widetilde{\nu}_R \bar{\psi}_N \psi_N$ can be defined, thus the total spin-independent sneutrino-nucleon scattering cross section\footnote{The sneutrino is a scalar field, then there is no axial-vector coupling in the effective Lagrangian, i.e. sneutrino DM results in vanishing spin-dependent cross section.} is given by~\cite{Han:1997wn,Cerdeno2:2009}
\begin{equation}
    \sigma^{SI}_{\widetilde{\nu}_R-N}\; = \; \frac{\mu_{red}^2}{\pi m_{\widetilde{\nu}_R}^2} g_N^2,
\end{equation}
where $\mu_{red} = m_N m_{\widetilde{\nu}_R} / (m_N + m_{\widetilde{\nu}_R})$ is the reduced mass of the nucleon with mass $m_N$. The effective coefficient results~\cite{Han:1997wn,Cerdeno2:2009}
\begin{align}
    g_N \; &= \; m_N \sum^3_{i=1} \frac{C_{\widetilde{\nu}_R \widetilde{\nu}_R h_i} C_{N N h_i}}{m_{h_i}^2} \nonumber \\
    &= \; m_N \sum^3_{i=1} \frac{C_{\widetilde{\nu}_R \widetilde{\nu}_R h_i} }{m_{h_i}^2} \left( \sum_{q_j=u,c,t} \frac{f_{q_j}^{(N)} \; Y_{q_j}}{m_{q_j}} \; S_{iH_u} + \sum_{q_j=d,s,b} \frac{f_{q_j}^{(N)} \; Y_{q_j}}{m_{q_j}} \; S_{iH_d} \right) \nonumber\\
    &= \; m_N \sum^3_{i=1} \frac{C_{\widetilde{\nu}_R \widetilde{\nu}_R h_i} }{m_{h_i}^2} \left( \frac{\sqrt{2} F_u^{(N)}}{v_u} S_{iH_u} + \frac{\sqrt{2} F_d^{(N)}}{v_d} S_{iH_d} \right),
    \label{effcoupling}
\end{align}
where $S_{iH_u}$ and $S_{iH_d}$ are the elements of the matrices defined in Eq.~(\ref{cphiggsmatrices}) that diagonalize the CP-even Higgs mass matrix. $C_{N N h_i}$ involves the Yukawa couplings of the Higgs bosons with the constituents of the nucleon $N$ represented by its form factors $F_u^{(N)}$ and $F_d^{(N)}$ subject to considerable uncertainties, with $f^{(N)}_q \; = \; m_N^{-1} \; \langle  N  | \; m_q  q  \bar{q} \; | N \rangle$. Finally, $C_{\widetilde{\nu}_R \widetilde{\nu}_R h_i}$ determines the sneutrino-sneutrino-Higgs coupling strength involving the terms $\lambda_N \widetilde{N} \widetilde{N} S$, $\lambda S H_u H_d$, $\kappa S^3$, and $A_{\lambda_N} \lambda_N \widetilde{N} \widetilde{N} S$. It is defined as~\cite{Cerdeno2:2009}
\begin{align}
    C_{\widetilde{\nu}_R \widetilde{\nu}_R h_i} = \frac{\lambda \lambda_N}{\sqrt{2}} \left(  v_u S_{iH_d} \; + \; v_d S_{iH_u}  \right) \; + \; \left[  (4 \lambda_N^2 \; + \; 2 \kappa \lambda_N ) \; \frac{\mu_{eff}}{\lambda} \; + \; \frac{A_{\lambda_N} \lambda_N}{\sqrt{2}} \right] S_{iS},
\end{align}
the first two terms represent the coupling with the Higgs doublets and the last terms with the singlet.

To estimate the order of magnitude of the spin-independent sneutrino-proton cross section let us consider that the only relevant scattering proceeds by exchanging the lightest Higgs boson $h_1$ that is SM-like, then
\begin{equation}
    \sigma^{SI}_{\widetilde{\nu}_R-p}\; = \; \frac{(F_u^{(p)})^2}{\pi} \frac{m_p^4}{m_{\widetilde{\nu}_R}^2m_{h_1}^4} \left( \frac{\lambda \; \lambda_N}{\tan \beta} \right)^2,
    \label{sigmaSIapprox1}
\end{equation}
where we have approximated $\mu_{red} \simeq m_p$ for $m_p \ll m_{\widetilde{\nu}_R}$, with $m_p$ the proton mass. If we take $F_u^{(p)}\simeq 0.15$ as default value used by 
\texttt{MicrOmegas}~\cite{micromegascite1,micromegascite2,micromegascite3} we get
\begin{equation}
    \sigma^{SI}_{\widetilde{\nu}_R-p}\; = \; 1.14\times 10^{-48} \text{cm}^{2} \left( \frac{100 \text{ GeV}}{m_{\widetilde{\nu}_R}} \right)^2 \left( \frac{\lambda}{0.1} \right)^2 \left( \frac{\lambda_N}{0.1} \right)^2 \left( \frac{10}{ \tan \beta} \right)^2.
\end{equation}

Eq.~(\ref{sigmaSIapprox1}) can be used as a first order approximation as most solutions found present a SM-like Higgs as the lightest CP-even state, and a Higgs boson with $H_d^0$ dominant component whose contribution to the cross section is suppressed by its large mass. Regarding the scattering with a nucleon via a $t$-channel exchange of a singlet dominated scalar, its contribution is suppressed by the mixing with the doublets, as can be seen from Eq.~(\ref{effcoupling}).

From the above discussion, the sneutrino direct detection depends on both the Higgs sector, mainly through $\lambda$ and $\tan \beta$, and the sneutrino parameters $\lambda_N$ and $m_{\widetilde{\nu}_R}$, involved in Eq.~(\ref{RHsneutrinomassAprox}).

\section{Scan details and experimental constraints}
\label{sec:scandetails}

\subsection{Sampling setup and strategy}
\label{samplinsetup}

To obtain representative solutions for different DM candidates, and explore the relevant parameter space of the model, we carried out a series of scans. 
To find regions compatible with a given experimental data, we used a likelihood data-driven method employing the \texttt{Multinest}~\cite{Feroz:2008xx} algorithm as optimizer\footnote{The main focus of this work is to present characteristic features of the model regarding the DM candidates. We do not aim to perform an statistical interpretation as done in Ref.\cite{Cao:2019}.}. 

We used the \texttt{Mathematica} package \texttt{SARAH}~\cite{sarahcite1,sarahcite2,sarahcite3} to build the model, and the code \texttt{SPheno}~\cite{sphenocite1,sphenocite2} to generate the particle spectrum, branching ratios and decay rates. Each point is required not to have tachyonic eigenstates and we only select models that have the lightest neutralino or the lightest RH sneutrino as LSP. Then, we compute the likelihood associated to each experimental data set.

\texttt{MicrOmegas}~\cite{micromegascite1,micromegascite2,micromegascite3} is used to compute the DM relic density, the present annihilation cross section ($\langle \sigma_{DM} v\rangle$), and the spin-dependent and spin-independent WIMP-nucleon scattering cross sections ($\sigma^{\text{SD}}_{DM-p}$ and $\sigma^{\text{SI}}_{DM-p}$), assuming that the mentioned candidates are the sole DM candidate in the Universe. However, we do not assume that the thermal relic density saturates the Planck value in order to allow for the possibility of multicomponent DM. For example, axions might make up a substantial amount of DM, or even gravitinos coexisting with RH sneutrinos are possible for some parameter regions.

For the DM annihilation spectrum we consider constraints on DM indirect detection searches obtained from the observation of dwarf galaxies, using \Fermi-LAT collaboration data~\cite{Ackermann:2015dwarfs}, and the observation of the Galactic center, taken from H.E.S.S. collaboration analysis~\cite{HESS:2016}. Both data sets were implemented as hard cuts on each of the reported channels. Other limits, for example on line signals, are weaker for the neutralino and RH sneutrino signatures. 

The constraints on the WIMP-nucleon scattering cross sections from DM direct detection experiments (XENON1T~\cite{xenon1tcite1,xenon1tcite2} and PICO-60~\cite{picocite1,picocite2}) are computed using \texttt{DDCalc}~\cite{ddcalccite}. We required that the p-value reported by \texttt{DDCalc} be larger than 5\% during the scan. Later we apply XENON1T and PICO-60 central values as hard-cuts in our analysis.
DM direct detection limits are rescaled by $r_{DM}=\frac{\Omega_{DM}h^2}{\Omega_{cdm}^{\text{Planck}}h^2}$, and the indirect detection constraints by $r_{DM}^2$, if the thermal relic density is less than the observed value.

\texttt{HiggsBounds}~\cite{HBcite1,HBcite2,HBcite3} is used to determine whether the SUSY models satisfy LEP, Tevatron and LHC Higgs constraints. Negative searches of Higgs-like signals were transformed into exclusions limits, and given a parameter point with its theoretical prediction in the Higgs sector, \texttt{HiggsBounds} indicates if the parameter set is allowed or not at 95\% confidence level with an step likelihood function (i.e. allowed: 1, excluded: 0). In order to assess if the predicted Higgs sector  reproduces the signal observed by ATLAS and CMS complemented with Tevatron data, \texttt{HiggsSignals}~\cite{HScite1,HScite2,HScite3} is used to quantitatively determine with a $\chi^2$ measure the compatibility of the NMSSM prediction with the measured signal strength and mass. We required that the p-value reported by \texttt{HiggsSignals} be larger than 5\%.

\subsection{Flavor and other collider constraints}
\label{sec:flavorandLHC}



In addition to the experimental constraints mentioned in the sampling setup, we considered several flavor and SUSY searches that we discuss below. A summary of all the constraints adopted for our model can be seen in Table~\ref{TableConstraints}.  \\

\begin{table}
\begin{small}
\begin{center}
\begin{tabular}{|c|c|}

        \hline
        \textbf{Constraint} & \textbf{} \\
        \hline
        \hline
            DM relic density ($\Omega_{cdm}^{\text{Planck}}h^2$) & $0.1198 \pm 0.0012$~\cite{Planck:2018VI} \\
            SI cross section ($\sigma^{\text{SD}}_{DM-p}$) & XENON1T~\cite{xenon1tcite1,xenon1tcite2} \\
            SD cross section ($\sigma^{\text{SI}}_{DM-p}$) & PICO60~\cite{picocite1,picocite2} \\
            Annihilation cross section ($\langle \sigma_{DM} v\rangle$) & \Fermi-LAT~\cite{Ackermann:2015dwarfs} and H.E.S.S.~\cite{HESS:2016} \\
            Higgs constraints & LEP, Tevatron, and LHC (\texttt{HiggsBounds}~\cite{HBcite1,HBcite2,HBcite3}) \\
            Higgs signal & LHC and Tevatron (\texttt{HiggsSignals}~\cite{HScite1,HScite2,HScite3}) \\
            $BR(b \rightarrow s \gamma)$ & $(3.27 \pm 0.14) \times 10^{-4}$~\cite{Misiak:2017} \\
            $BR(B_s \rightarrow \mu^+ \mu^-)$ & $(2.8^{+0.8}_{-0.7}) \times 10^{-9}$~\cite{ATLASBflavor:2019} \\
            $BR(\mu \rightarrow e \gamma)$ & $< 4.2 \times 10^{-13}$~\cite{MEGmuegamma:2016} \\
            $BR(\mu \rightarrow eee)$ & $< 1.0 \times 10^{-12}$~\cite{SINDRUMmueee:1988} \\
            Light stops and sbottoms & LHC~\cite{ATLASdm:2018,Aaboud:2017aeu} \\
            R-hadrons (long-lived colored particles) & LHC~\cite{ATLASllcharged:2019} \\
            2 and 3 Leptons + missing $E_T$ & LHC~\cite{Aad:2014vma,Aaboud:2018jiw,Aad:2019vnb} \\
            Chargino masses & $m_{\tilde{\chi}^{\pm}_1} > 103.5$ GeV~\cite{Zyla:2020zbs} \\
        \hline
\end{tabular}
\end{center}
\end{small}
  \caption{Constraints that have been applied to our model set (see text for details).}
  \label{TableConstraints}
\end{table}

\textbf{Flavor:} We take into account current constraints on some flavor observables calculated with \texttt{SPheno}. $b \rightarrow s \gamma$ is a flavour changing neutral current (FCNC) process forbidden at tree level in the SM. However, it occurs at leading order through loop diagrams and becomes potentially sensitive to new physics.
Similarly, $B_s \rightarrow \mu^+ \mu^-$ is also forbidden at tree level in the SM but occurs radiatively. We use the following experimental determinations~\cite{Misiak:2017,ATLASBflavor:2019}: 
\bea
BR(b \rightarrow s \gamma) &=& (3.27 \pm 0.14) \times 10^{-4}, \\
BR(B_s \rightarrow \mu^+ \mu^-) &=& (2.8^{+0.8}_{-0.7}) \times 10^{-9}.
\eea
For $BR(b \rightarrow s \gamma)$ we considered the calculated average in Ref.~\cite{Misiak:2017} using the experimental values~\cite{BABAR:2008,BABAR:2012,BABAR:2012b,Belle:2015,Belle:2016}. For $BR(B_s \rightarrow \mu^+ \mu^-)$ we considered the ATLAS Collaboration determination~\cite{ATLASBflavor:2019}, which is in agreement with the LHCb measurement based on 8 TeV data~\cite{LHCbflavor:2017} and a former statistical combination of CMS and LHCb measurements with 7 and 8 TeV data~\cite{CMSLHCbflavor:2015}. We have also considered the theoretical uncertainties for each observable as 10\% of the corresponding best fit value. 
We do not include constraints as the muon anomalous magnetic moment, or $B_d \rightarrow \mu^+ \mu^-$~\cite{LHCbflavor:2017}, since we are not trying to solve any possible discrepancy with respect to the SM predictions.

The SM allows charged lepton flavour violating (LFV) processes with only extremely small branching ratios ($< 10^{-50}$) even taking into account neutrino mass differences and mixing angles. Such decays free from SM background are very sensitive to new physics, in particular to SUSY models. We considered the LFV constraint from $BR(\mu \rightarrow e \gamma) < 4.2 \times 10^{-13}$~\cite{MEGmuegamma:2016} at 90\% C.L., and $BR(\mu \rightarrow eee) < 1.0 \times 10^{-12}$~\cite{SINDRUMmueee:1988} at 90\% C.L.\\

\textbf{Light stops and sbottoms:} Due to our choice of free parameters and fixed values (see the next subsection), the lightest squarks are mainly stops and sbottoms, therefore the constraints we consider focus on these particles. Some neutralinos, especially Bino dominated, can coannihilate with a colored particles to obtain an allowed amount of relic density. The two-body decay channel $\tilde{t}_1 \rightarrow t \, \chi_1^0$ and three-body decay channel $\tilde{t}_1 \rightarrow b \, W \, \chi_1^0$ are kinematically forbidden when the lightest stop is nearly degenerate with the neutralino, which is needed to get an efficient coannihilation mechanism. 
Hence, we also take into account current constraints for a compressed mass spectrum. The dominant light stop decay would be via flavor-changing neutral current (FCNC) two-body decay channel $\tilde{t}_1 \rightarrow c \, \chi_1^0$, and a contribution given by the four-body decay channel $\tilde{t}_1 \rightarrow b \, f \, \bar{f}' \, \chi_1^0$. If the lightest colored particle is a sbottom, then $\tilde{b}_1 \rightarrow b \, \chi_1^0$ is relevant. The exclusion limits are given in term of the stop (or sbottom) and neutralino masses~\cite{ATLASdm:2018,Aaboud:2017aeu}.

RH sneutrinos can also achieve an allowed relic abundance through coannihilations with squarks. Due to the low interaction rate between the RH sneutrino and the lightest squark, along with the small mass splitting to get efficient coannihilation, stops and sbottoms would have decay lengths larger than 100 kilometers. For these scenarios, constraints on long-lived colored particles at the LHC, called R-hadrons, have to be applied. Hence, we consider ATLAS constraints with $\sqrt{s}=13$ TeV and $L=36.1$ fb$^{-1}$, imposing $m_{\tilde{t}}>1340$ GeV and $m_{\tilde{b}}>1250$ GeV~\cite{ATLASllcharged:2019} for RH sneutrinos LSP with stop or sbottom NLSP, respectively.\\

\textbf{Leptons + missing $E_T$ final states:} LHC searches for electroweak production of charginos and sleptons decaying into final states with 2 and 3 leptons plus missing transverse energy are relevant, in particular for Bino dominated $\chi_1^0$ and Wino dominated $\chi_2^0$ and $\chi_1^{\pm}$~\cite{Aad:2014vma,Aaboud:2018jiw,Aad:2019vnb}. \\



\textbf{Charginos:} We apply a cut for the chargino masses ($m_{\tilde{\chi}^{\pm}_1} > 103.5$ GeV) following LEP searches~\cite{Zyla:2020zbs}. Finally, we would like to mention that coannihilations between the lightest neutralino LSP and the lightest chargino NLSP, may imply disappearing-track signatures in pp collisions due to long-lived charginos, if their masses are nearly degenerate. However, extremely pure Higgsinos or Winos and neutrino-chargino mass difference of $\approx 160$ MeV~\cite{ATLASllcharginos:2018} are needed. We do not have any point in our scan fulfilling those conditions, hence we do not have exclusions from the latter restrictions.


\subsection{Input parameters}
\label{sec:inputparameters}


Following the considerations made in previous sections, in the sfermion sector we fix the dimensional parameters that are not especially relevant to our analysis, $m^2_{\tilde{e}_i}=m^2_{\tilde{L}_i}=m^2_{\tilde{d}_i}=2.25\times 10^{6}$ GeV$^2$ with $i=1,2,3$, and $m^2_{\tilde{N}_i}=m^2_{\tilde{u}_i}=m^2_{\tilde{Q}_i}=2.25\times 10^{6}$ GeV$^2$ with $i=1,2$. The values taken are sufficiently large to be consistent with LHC sparticles searches.

We also set $T_d^{33}=T_{d_3}=256$ GeV, and $T_e^{33}=T_{e_3}=-98$ GeV taking into account the corresponding Yukawa couplings. The neutrino Yukawa couplings are only relevant to reproduce the neutrino mass pattern, hence they are set $Y_N^{i}=10^{-6}$ when the scan is focused on finding solutions with neutralino or RH sneutrino LSP. We consider vanishing $T_{N}^i=A_N^i \, Y_N^i$ as an approximation due to the small neutrino Yukawa couplings and that we take $A_{_N}^i\sim O(\text{GeV})$. 
To simplify the analysis, we set $\lambda_N^i=-0.5$, and $T_{\lambda_N}^{i} = 0$ for $i=1,2$, to obtain two families of heavy sneutrinos.

The gaugino sector is described by its soft-breaking masses $M_1$, $M_2$, and $M_3$; we fix the gluino mass parameter $M_3=3$ TeV to avoid LHC constraints on gluino strong production.

In the Higgs-scalar singlet sector the soft-breaking masses  are related with the vacuum expectation values (VEVs) by the  minimization conditions of the Higgs potential after electroweak symmetry breaking (EWSB). Then, it is conventional to take as free parameters (inputs) the following: the ratio of the Higgs VEVs $\tan \beta\equiv v_u/v_d$, $\lambda$, $\kappa$, the effective higgsino mass parameter $\mu_{eff}= \lambda \, v_s/ \sqrt{2}$, 
$T_{\lambda}=A_{\lambda} \, \lambda$ and $T_{\kappa} = A_{\kappa} \, \kappa$.

We are left with the following set of variables as independent parameters:
\begin{equation}
M_1, \hspace{0.3cm} M_2, \hspace{0.3cm} \tan \beta , \hspace{0.3cm} \mu_{eff}, \hspace{0.3cm} \lambda , \hspace{0.3cm} \kappa , \hspace{0.3cm} \lambda_N^3 , \hspace{0.3cm} T_{\lambda}, \hspace{0.3cm} T_{\kappa}, \hspace{0.3cm} T_{\lambda_N}^3 , \hspace{0.3cm} m^2_{\tilde{N}_3}, \hspace{0.3cm} m^2_{\tilde{u}_3}, \hspace{0.3cm} m^2_{\tilde{Q}_3}, \hspace{0.3cm} T_{u_3}.
\label{inputsvariables}
\end{equation}
We carried out a scan over these parameters within the ranges depicted in Table~\ref{scanparameters} using log priors (in logarithmic scale). The ranges were set taking into account the following considerations:

\begin{table}[t]
	\begin{minipage}{.57\linewidth}
    \begin{small}
      \centering
        \hspace{0.9cm}\begin{tabular}{|c|c|}
        \hline
        \textbf{Parameter} & \textbf{Range} \\
        \hline
        \hline
         & \\[-2ex]
            $M_1$ & (20, 3000) GeV \\[1ex]
            $M_2$ & (20, 3000) GeV \\[1ex]
            $\mu_{eff}$ & (100, 5000) GeV \\[1ex]
            $\tan \beta$ & (2, 50) \\[1ex]
 & \\[-1ex]
            $\lambda$ & (0.001, 0.8) \\[1ex]
            $\kappa$ & (0.001, 0.8) \\[1ex]
            $\lambda_N^3$ & (-0.4, -0.001) \\[1ex]
            $T_{\lambda}$ & (0.001, 600) GeV \\[1ex]
            $T_{\kappa}$ & (-30, -0.001) GeV \\[1ex]
            $T_{\lambda_N}^3$ & (-1100, -0.001) GeV \\[1ex]
 & \\[-1ex]
            $m^2_{\tilde{N}_3}$ & ($10$, $2.5\times10^{6}$) GeV$^2$ \\[1ex]
            $m^2_{\tilde{u}_3}$ & ($2.5\times10^{5}$, $4\times10^{6}$) GeV$^2$ \\[1ex]
            $m^2_{\tilde{Q}_3}$ & ($2.5\times10^{5}$, $4\times10^{6}$) GeV$^2$ \\[1ex]
            $T_{u_3}$ & (700, 10000) GeV \\[1ex]
        \hline
        \end{tabular}
    \end{small}
    \end{minipage}%
    \begin{minipage}{.5\linewidth}
    \begin{small}
      \centering
        \begin{tabular}{|c|c|}
        \hline
        \textbf{Parameter} & \textbf{Fixed value} \\
        \hline
        \hline
         & \\[-2ex]
            $M_3$ & 3000 GeV \\[1ex]
 & \\[-1ex]
            $\lambda_N^i$,\begin{footnotesize} $i=1,2$\end{footnotesize} & -0.5 \\[1ex]
            $T_{\lambda_N}^i$,\begin{footnotesize} $i=1,2$\end{footnotesize} & 0 GeV \\[1ex]
 & \\[-1ex]
            $m^2_{\tilde{N}_i}$,\begin{footnotesize} $i=1,2$\end{footnotesize} & $2.25\times10^{6}$ GeV$^2$ \\[1ex]
            $m^2_{\tilde{u}_i}$,\begin{footnotesize} $i=1,2$\end{footnotesize} & $2.25\times10^{6}$ GeV$^2$ \\[1ex]
            $m^2_{\tilde{Q}_i}$,\begin{footnotesize} $i=1,2$\end{footnotesize} & $2.25\times10^{6}$ GeV$^2$ \\[1ex]
            $m^2_{\tilde{d}_i}$,\begin{footnotesize} $i=1,2,3$\end{footnotesize} & $2.25\times10^{6}$ GeV$^2$ \\[1ex]
            $m^2_{\tilde{e}_i}$,\begin{footnotesize} $i=1,2,3$\end{footnotesize} & $2.25\times10^{6}$ GeV$^2$ \\[1ex]
            $m^2_{\tilde{L}_i}$,\begin{footnotesize} $i=1,2,3$\end{footnotesize} & $2.25\times10^{6}$ GeV$^2$ \\[1ex]
 & \\[-1ex]
            $T_{d_3}$ & 256 GeV \\[1ex]
            $T_{e_3}$ & -98 GeV \\[1ex]
            $T_{N}^i$,\begin{footnotesize} $i=1,2,3$\end{footnotesize} & 0 GeV \\[1ex]
        \hline
        \end{tabular}
    \end{small}
    \end{minipage} 
  \caption{Sampling ranges and fixed parameters used in our scan.}
  \label{scanparameters}
\end{table}

\begin{itemize}
\item The range of $\tan \beta$, $T_{u_3}$ and the soft squark masses of the 3rd generation, are helpful to reproduce the correct SM-like Higgs mass.
\item The upper bounds of $\lambda$, $\kappa$, $\lambda_N$ and $\tan \beta$ are set to satisfy perturbativity of the theory up to Planck scale.
\item LEP searches for chargino and neutralino requires $m_{\tilde{\chi}^{\pm}_1} > 103.5$ GeV, then $\mu_{eff} > 100$ GeV. We also expect $\mu_{eff} \simeq O(100)$ GeV because it is directly related to $m_Z$, but we allow values of $O(\text{TeV})$ to be general.
\item The range of $M_1$, $M_2$, $\mu_{eff}$, $\lambda$ and $\kappa$ allow us to find neutralinos with different compositions.
\item The signs of $\lambda_N$, $T_{\lambda_N}$ and $\kappa$, are set to obtain a CP-even state as the lightest RH sneutrino (see Eq.~\ref{RHsneutrinomassAprox}).
\item The sign of $T_{\kappa}$ is set to allow light CP-odd Higgs state with dominant singlet contribution.
\item The range and fixed values of $\lambda_N^i$, $T_{\lambda_N}^i$ and $m_{\tilde{N}_i}^2$, allow us to find one family of light RH sneutrinos and two heavy ones.
\end{itemize}

\section{Results}
\label{sec:scanresults}

\subsection{Relic density}
\label{sec:relicdensity}

\begin{figure}[t!]
\begin{center}
 \begin{tabular}{c}
 \hspace*{-10mm}
 \epsfig{file=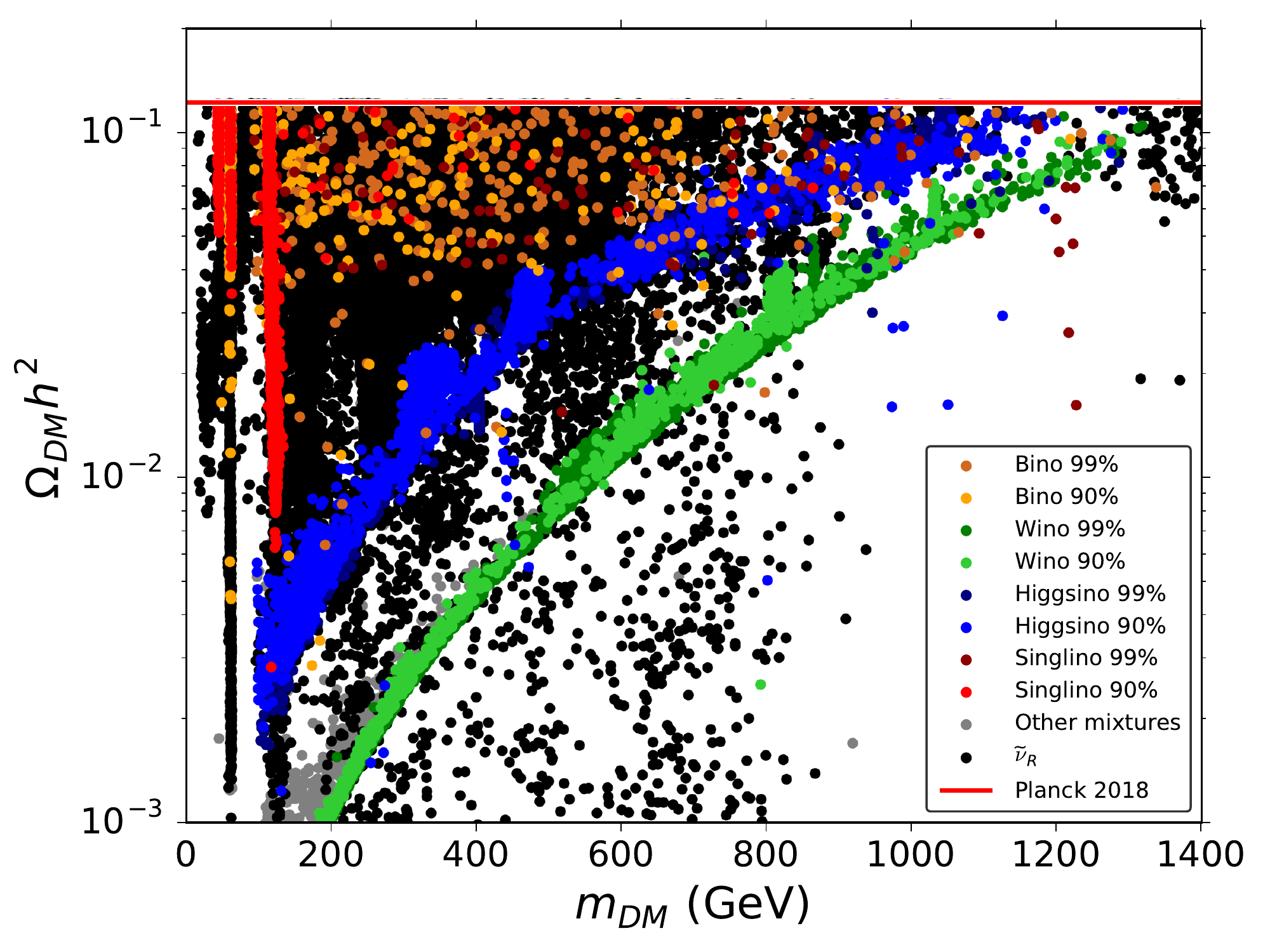,height=9cm} 
    \end{tabular}
    \captions{Relic density versus LSP mass for the parameter points that fulfill all the constraints considered in this work. The color coding represents the LSP identity, and in the case of neutralino, the dominant composition as labeled. The red solid line corresponds to the amount of DM measured by the Planck Collaboration.
}
    \label{relicscan}
\end{center}
\end{figure}


In Fig.~\ref{relicscan} we show the relic density as a function of the DM candidate mass for the parameter points that fulfill the constraints considered in this work. The results correspond to the broad range of input parameters shown in Table~\ref{scanparameters}, with additional explorations in several input parameter subranges. The identity of the LSP, and dominant composition for the neutralino\footnote{Through out the rest of the text we will refer to the lightest neutralino as neutralino for simplicity, unless otherwise specified.} case, is coded in color as indicated on the figure (we define a pure neutralino if $|N_{1j}|^2>0.9$, with dominant component $j$ defined in Eq.~(\ref{diagneutralino})). Due to the fact that the off-diagonal elements of the neutralino and chargino mass matrices are at most $\sim m_W$ (see Eq.~(\ref{neutralinomatrix}) and (\ref{charginomatrix})), and the scan ranges chosen, neutralinos are typically pure electroweak eigenstate.

For the neutralino DM case, the mass lower limit is established by a combination of LEP and relic density constraint, while the upper limit due to the input parameter ranges. One of the most stringent limits that constraints Bino and Singlino dominated neutralinos is the Planck upper limit on cold DM abundance, since models with the mentioned candidates tend to produce too high relic density. 
However, points with this kind of neutralinos and $\Omega_{DM}h^2=\Omega_{cdm}^{\text{Planck}}h^2$ can be found in the entire mass range. On the other hand, solutions that saturate the relic density with Higgsinos or Winos are restricted to $1000\sim 1400$ GeV DM masses.

\begin{figure}[t!]
\begin{center}
 \begin{tabular}{c}
 \hspace*{-10mm}
 \epsfig{file=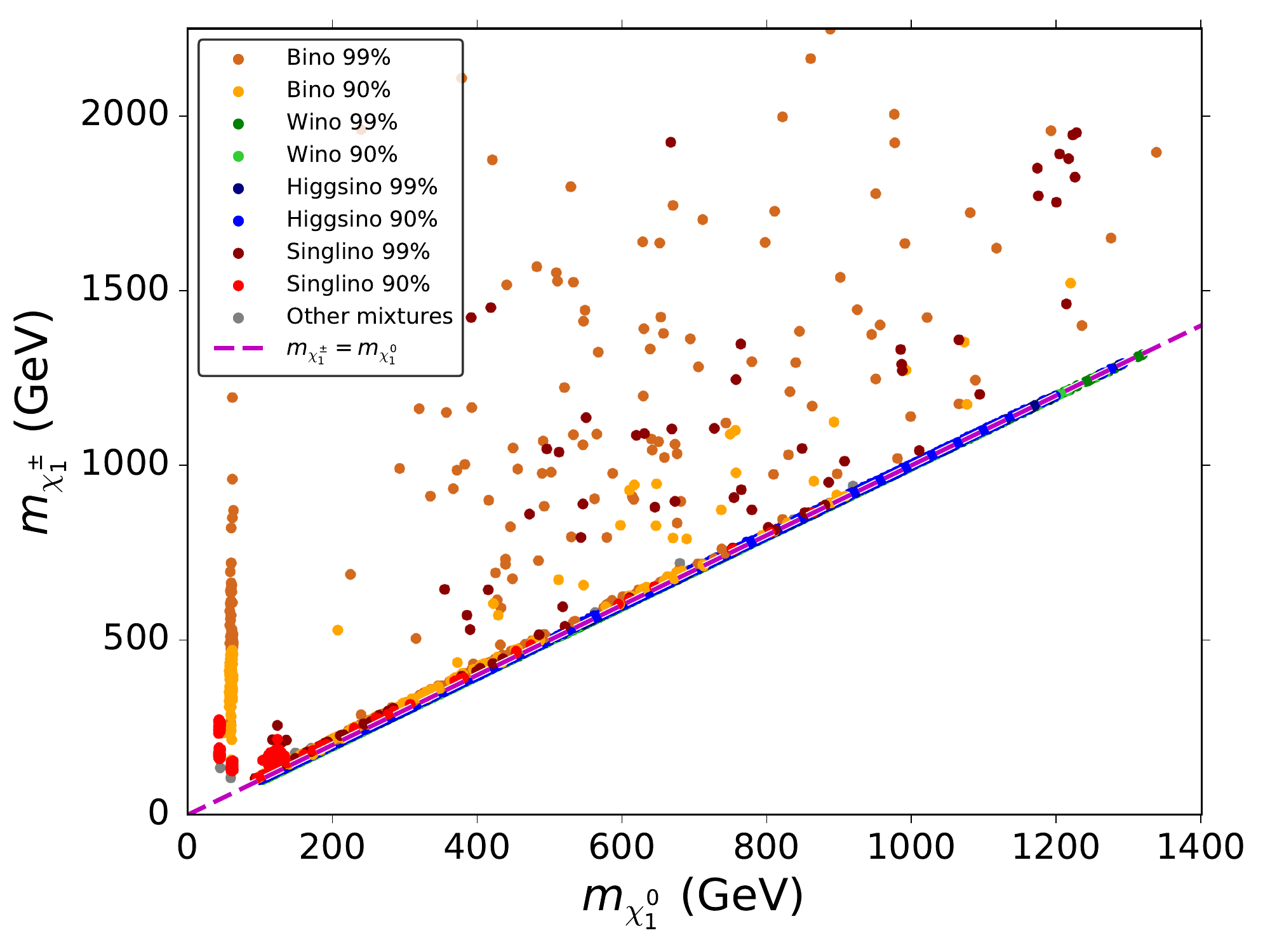,height=9cm} 
    \end{tabular}
    \captions{Lightest chargino vs lightest neutralino. Solutions with RH sneutrino LSP are not included. Almost all Higgsino and Wino (below the blue points) dominated neutralinos coannihilate with charginos to obtain an allowed relic density. Low mass Bino and Singlino annihilation mechanisms employ resonances.
}
    \label{paramneutralinos}
\end{center}
\end{figure}

To briefly analyse the channels used by different types of neutralinos, in Fig.~\ref{paramneutralinos} we show the lightest chargino vs the neutralino mass, without including the points with RH sneutrino as DM candidate. We can clearly see that for almost all Higgsino and Wino dominated neutralinos, coannihilations with charginos are important to obtain an allowed relic abundance. Notice that Wino-like neutralinos in green lie below the blue points. Singlinos with non negligible Higgsino contribution also employ this mechanism. For low masses, resonant conditions through $Z$ and $h_i$ are dominant, and depicted as vertical narrow strips (see also Fig.~\ref{relicscan}). For higher masses, coannihilation channels with squarks are relevant for Bino and Singlino neutralinos.

Two relevant parameter in the neutralino sector are $\lambda$ and $\kappa$, related to the Higgsino and Singlino masses, respectively (see Eq.~(\ref{neutralinomatrix})). As expected, $\chi_1^0$ can be Higgsino dominated for $\lambda < 2 \, \kappa$, and Singlino dominated if $\lambda > 2 \, \kappa$. Notice that the gauginos are not directly coupled with the Singlinos, hence Bino and Wino dominated neutralinos can be achieved for any relation between $\lambda$ and $\kappa$. 
The mentioned parameters will be relevant later when we discuss RH sneutrino DM SI cross section, since $\lambda$ is involved in Eq.~(\ref{sigmaSIapprox1}). This will be particularly important for RH sneutrino-Higgsino  coannihilations.

Next we will focus on the sneutrino sector. Similarly to the case of Bino and Singlino dominated neutralinos, the small coupling of the RH sneutrino with the rest of the particles results in a very high RH sneutrino relic abundance. However, as seen in Subsec.~\ref{sec:sneutrinoDM}, there are three main mechanisms to efficiently annihilate the RH sneutrinos:
the resonant sneutrino mass condition $m_{\tilde{\nu}_R} \simeq m_{h_i}/2$, annihilations via direct quartic couplings, and the coannihilation condition where the mass splitting between a RH sneutrino and a second sparticle is small.

\begin{figure}[t!]
\begin{center}
 \begin{tabular}{cc}
 \hspace*{-14mm}
 \epsfig{file=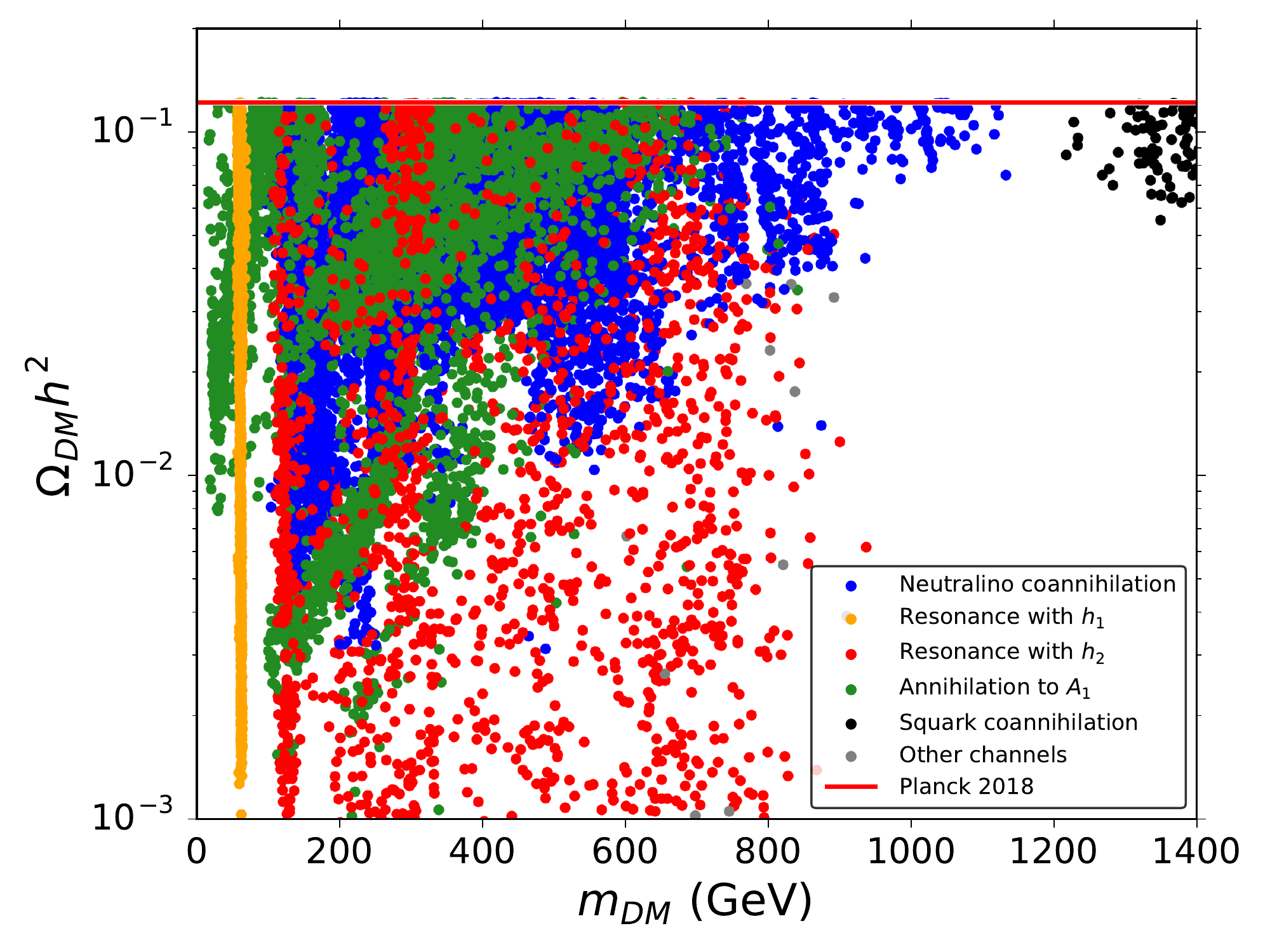,height=7cm} 
        \hspace*{-6mm}\epsfig{file=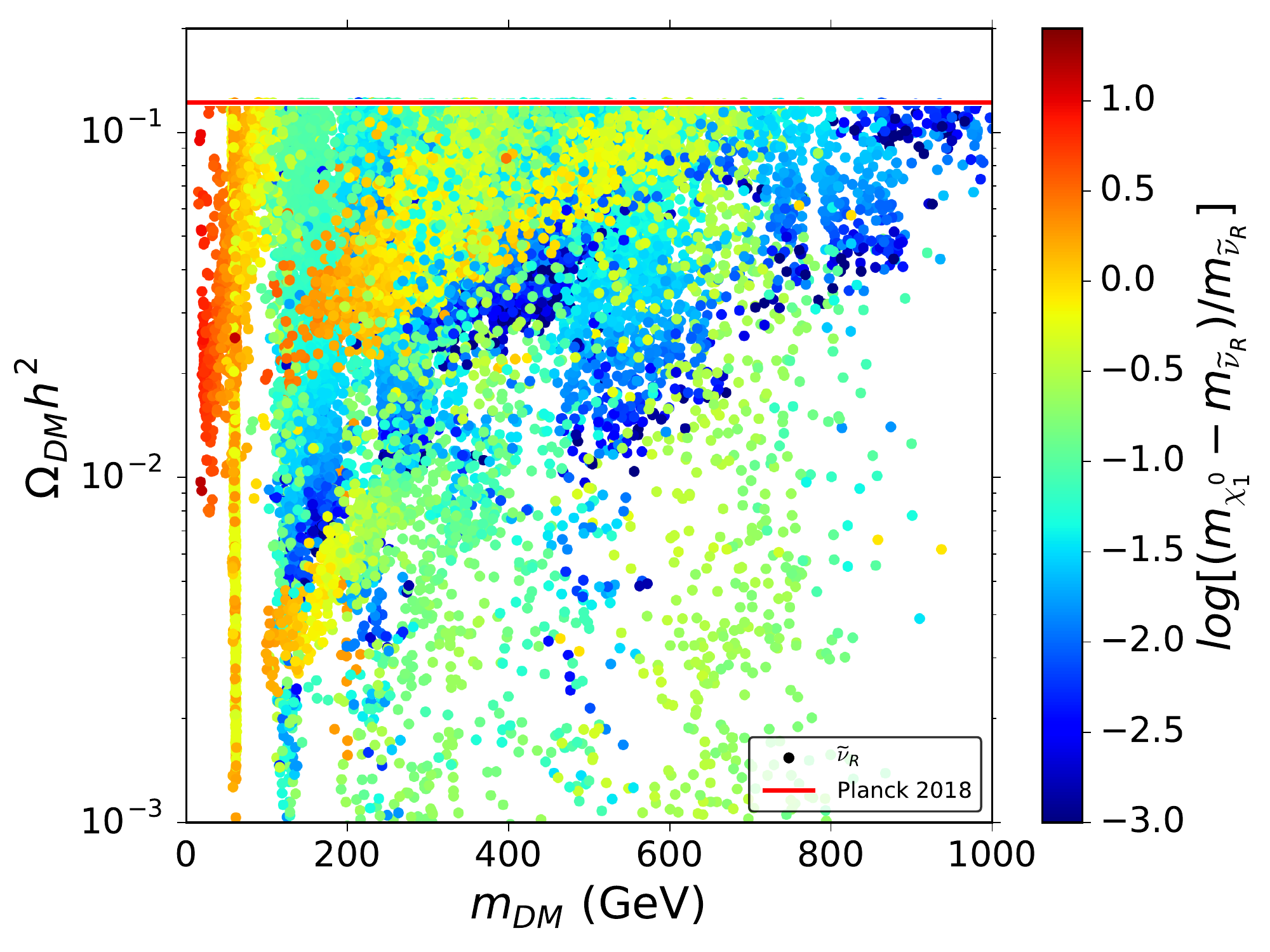,height=7cm}   
    \end{tabular}
    \captions{Relic density versus LSP mass, only for points with RH sneutrino DM. The color coding on the left panel represents the main channels used by RH sneutrinos to obtain an allowed relic density. The color coding on the right panel shows the weighted mass difference between neutralino and RH sneutrino. Points towards the blue tone correspond mainly to coaniquilations with neutrinos. The red solid line corresponds to the amount of DM measured by the Planck Collaboration.
}
    \label{relicscanRHsneutrinos}
\end{center}
\end{figure}

In Fig.~\ref{relicscanRHsneutrinos} we show only the points with RH sneutrino as DM candidate. The left panel depicts the main channels used by RH sneutrinos to obtain an allowed relic density. For $m_{\tilde{\nu}_R} \lsim 100$ GeV, the dominant processes are resonances with the lightest CP-even scalar (orange) and annihilations via direct quartic couplings to the lightest pseudo-scalar (green), with almost no coannihilations due to collider constraints on charginos. On the other hand, for $m_{\tilde{\nu}_R} \gsim 100$ GeV the dominant mechanisms are coannihilations with neutralinos (blue). However, very important contributions come from resonances with the second lightest CP-even scalar (red), and annihilations via direct quartic couplings. The former mechanism is especially important to obtain low relic densities and can be the only channel present for low enough abundances. The latter mechanism will be relevant for direct detection experiments, being the dominant channel to yield signals in the ballpark of next generation instruments. Coannihilations with stops are present for $m_{\tilde{\nu}_R} \gsim 1200$ GeV with extremely low sneutrino-nucleon scattering cross section. We would like to remark that in each mass region, solutions for the mentioned channels with $\Omega_{DM}h^2=\Omega_{cdm}^{\text{Planck}}h^2$ can be found.

An important requirement for coannihilation solutions is that the coannihilation partner should have efficient annihilation channels. If we decouple the RH sneutrinos, the relic abundance of the remaining LSP has to be within experimental constraints. Also, the relic density that RH sneutrinos can achieve with the mentioned mechanism is bounded from below by the relic density that would have obtained its coannihilation partner if it were the LSP.

The color coding of the right panel of Fig.~\ref{relicscanRHsneutrinos} represents the value of $log [ (m_{\chi_1^0} - m_{\tilde{\nu}_R})/m_{\tilde{\nu}_R} ]$. Coannihilations with neutralinos are important for a mass splitting $\lesssim 10 \%$, i.e. when $log [ (m_{\chi_1^0} - m_{\tilde{\nu}_R})/m_{\tilde{\nu}_R} ] \lesssim -1$. 
In Fig.~\ref{relicscan}, we can see that Higgsinos and Winos LSP are arranged in two easily identifiable curves. In the right panel of Fig.~\ref{relicscanRHsneutrinos}, points in dark blue, i.e. RH sneutrinos LSP with the lowest mass splitting, lie in the regions drawn by the mentioned curves. Furthermore, comparing with the left panel of Fig.~\ref{relicscanRHsneutrinos}, we can also identify these curves as the relic density lower limit of RH sneutrino DM with Higgsino or Wino coannihilations.

To summarize, the parameter space of RH sneutrino-neutralino coannihilations lies within 
\begin{equation}
    m_{\tilde{\nu}_R}^{LSP} < m_{\chi_1^0}^{LSP} \hspace{1cm} \text{and} \hspace{1cm} \Omega_{\chi_1^0}^{LSP}h^2 < \Omega_{\tilde{\nu}_R}^{LSP}h^2 < \Omega_{cdm}^{\text{Planck}}h^2,
\end{equation}where the subscript $\chi_1^0$ ($\tilde{\nu}_R$) with the superscript $LSP$ emphasizes whether the RH neutrino is (not) decoupled. Finally, notice that for a fixed DM mass, Winos LSP can achieve lower relic densities, therefore the points using coannihilations with neutralinos in the region below the Higgsino curve are dominated by Wino-like neutralinos.

We should highlight here that, as expected, it is much easier for \texttt{Multinest} to find neutralinos rather than RH sneutrinos as DM candidate fulfilling all the constraints imposed (if we run the scan with the input parameter broad range shown in Table~\ref{scanparameters} we get about $O(20)$ candidate points with neutralino LSP per point with sneutrino LSP). The solutions involving coannihilation channels with RH sneutrino are usually found after \texttt{Multinest} finds a viable neutralino. Then, to minimize the likelihood, the code scans a similar parameter region and a viable RH sneutrino could be achieved as a result of this process.

The above discussion can be used as a hint to search for RH sneutrino with the correct relic density. If a parameter point with neutralino DM is viable, varying $m_{\tilde{N}}$, $\lambda_N$, $T_{\lambda_N}$, and $\kappa$ according to Eq.~(\ref{RHsneutrinomassAprox}), we can get different values of $m_{\tilde{\nu}_R}$ while leaving approximately unchanged the rest of the mass spectrum. Then, a RH sneutrino as DM candidate with mass $m_{\tilde{\nu}_{R}} \sim m_{\chi_1^0}$  could be found using a viable neutralino solution as seed and the coannihilation mechanism to obtain a correct thermal relic abundance. Fig.~\ref{relicscan} includes some examples where a deeper exploration of the parameter space was performed in this way.

\begin{figure}[t!]
\begin{center}
 \begin{tabular}{cc}
 \hspace*{-14mm}
 \epsfig{file=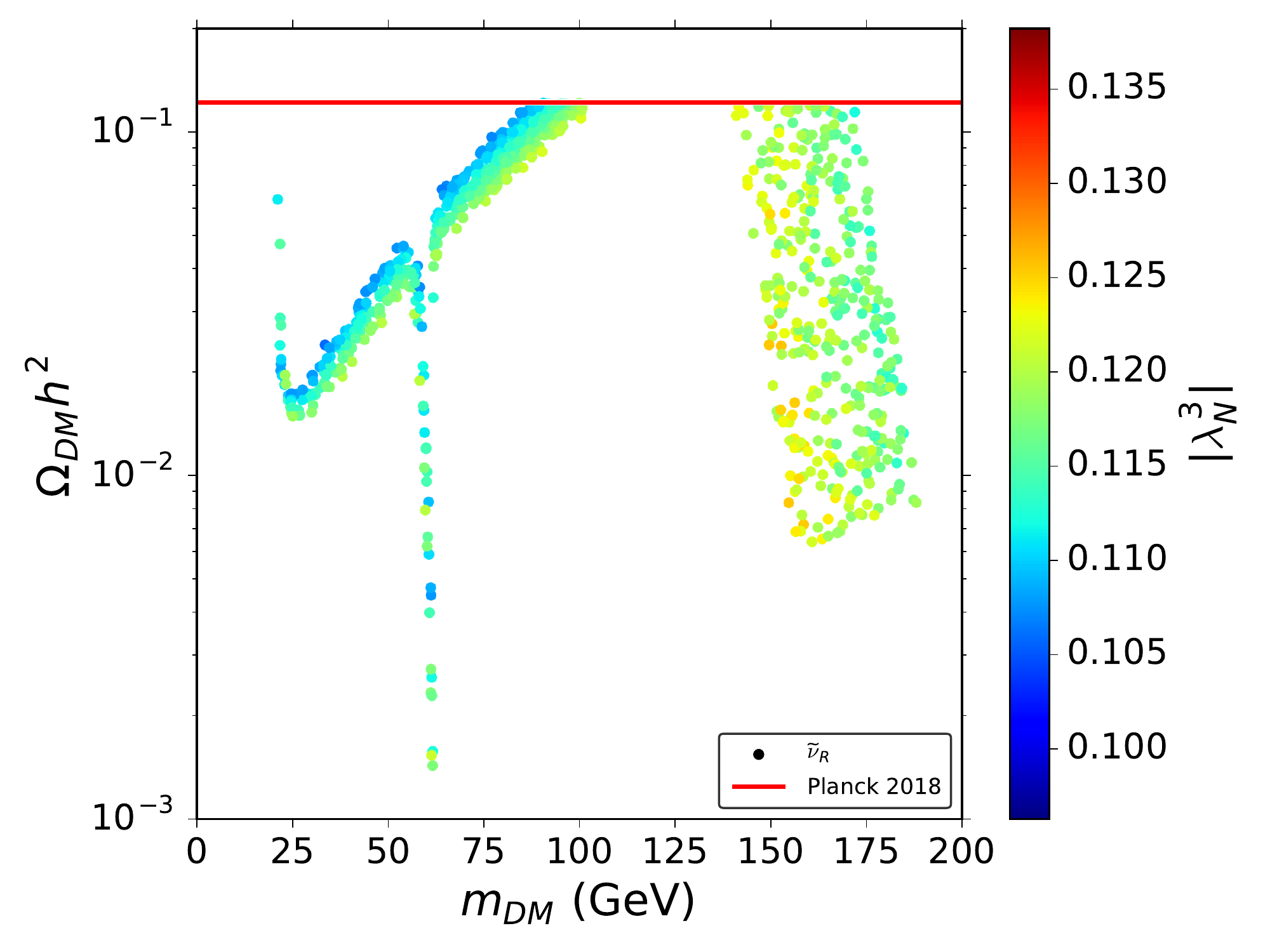,height=7cm} 
        \hspace*{-4mm}\epsfig{file=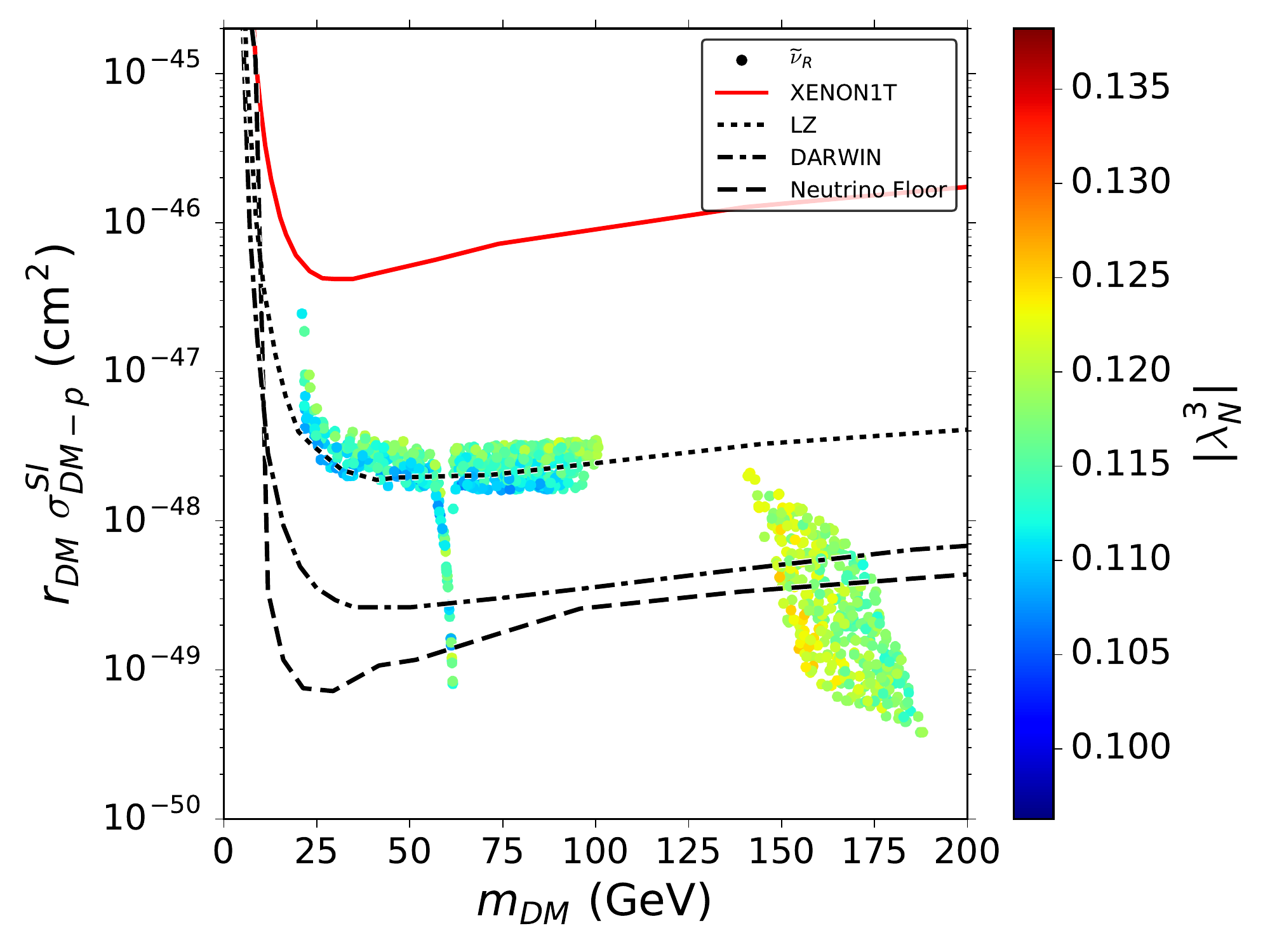,height=7cm}   
    \end{tabular}
    \captions{Relic density (left) and scaled spin-independent direct detection cross section (right) vs RH sneutrino mass for a scan in the ($\mu_{eff}$, $\lambda_N$, $T_{\lambda_N}$, $m^2_{\widetilde{N}_3}$) space (see text for details). Three annihilation channels can be identify: resonance with a SM-like Higgs and annihilation through scalar quartic coupling for $m_{DM} \lesssim 100$ GeV, and coannihilation with Higgsino dominated neutralinos for $m_{DM} \gtrsim 130$ GeV.
    }
    \label{scan63rrdata4}
\end{center}
\end{figure}

\begin{table}
\begin{small}
\begin{center}
\begin{tabular}{|c|c||c|c|}

        \hline
        \multicolumn{4}{|c|}{\textbf{Fixed Parameters}} \\
        \hline
        \hline
            $M_1$ & 1873.8 GeV & $m^2_{\tilde{u}_3}$ & 1.80$\times 10^{6}$ GeV$^2$ \\
            $M_2$ & 782.07 GeV & $m^2_{\tilde{Q}_3}$ & 1.16$\times 10^{6}$ GeV$^2$ \\
            $\tan \beta$ & 13.83 & $T_{u_3}$ & 1854.8 GeV \\
            $\lambda$ & 0.0750 & \hspace{0.7cm} $T_{\lambda}$ \hspace{0.7cm} & 194.0 GeV \\
            $\kappa$ & 0.0922 & $T_{\kappa}$ & -0.0195 GeV \\
            \hline
        \hline
        \multicolumn{4}{|c|}{\textbf{Scan Range}} \\
        \hline
        \hline
            $\lambda_N^3$ & [-0.126, -0.103] & $T_{\lambda_N}^3$ & [-38.8, -32.1] GeV \\
            $\mu_{eff}$ & [158.7, 192.1] GeV & $m^2_{\tilde{N}_3}$ & [3757.6, 4546.7] GeV$^2$ \\
        \hline
\end{tabular}
\end{center}
\end{small}
  \caption{Set of inputs for the scan shown in Fig.~\ref{scan63rrdata4}.}
  \label{tab:scan63}
\end{table}

To illustrate the rest of the main mechanisms to obtain solutions for different RH sneutrino masses, in Fig.~\ref{scan63rrdata4} we present the results of a dedicated scan. Three important channels can be easily identified: a resonance with a SM-like Higgs, annihilation through scalar quartic coupling, and coannihilation with Higgsino dominated neutralinos. The solutions with RH sneutrino presented in the figure were found considering a scan over only four parameters ($\mu_{eff}$, $\lambda_N$, $T_{\lambda_N}$, $m^2_{\widetilde{N}_3}$), whose range can be seen in Table~\ref{tab:scan63} together with the fixed value of rest of the parameters. This set of inputs results in a Higgsino dominated neutralino with $m_{\chi_1^0}\sim 156-190$ GeV, and a light pseudo-scalar with $m_{A_1}\sim 18-23$ GeV.

For $m_{\widetilde{\nu}_R}\lesssim 100$ GeV the first two mechanisms to achieve a correct amount of relic density can be noticed: a funnel condition for $m_{\widetilde{\nu}_R}\simeq m_{h_1}/2$ depicted as a vertical narrow strip, and annihilation through direct quartic coupling $\widetilde{\nu}_R \widetilde{\nu}_R \rightarrow A_1 A_1$. The latter coupling is determined by
\begin{equation}
    C_{\widetilde{\nu}_R \widetilde{\nu}_R A_1 A_1} \; = \; - \left(\lambda_N^2 - \frac{1}{2} \lambda_N \kappa \right),
\end{equation}
where we have assumed that the lightest pseudo-scalar is singlet dominated.  A more efficient annihilation and hence a lower relic density is achieved for increasing values of $\lambda_N$, as can be seen from the left panel of Fig.~\ref{scan63rrdata4}. In this case, allowed points with low mass sneutrinos can be obtained up to $m_{\widetilde{\nu}_R} \sim 20$ GeV when the four point interaction becomes kinematically forbidden. 

As discussed in Subsection~\ref{sec:RHsneutrinoDD} (see Eq.~(\ref{sigmaSIapprox1})), $\sigma^{SI}_{\widetilde{\nu}_R-p}$ increases for larger values of $\lambda_N$, i.e. larger sneutrino-sneutrino-Higgs coupling. This is shown on the right panel of Fig.~\ref{scan63rrdata4}, especially for $m_{\widetilde{\nu}_R} \lesssim 100$ GeV, where the solutions for this particular scan present low dispersion in $\Omega_{DM}h^2$ for each value of $m_{DM}$. Notice that for points using direct quartic coupling $r_{DM} \sim \Omega_{DM}h^2 \sim m_{DM}^{2}$ and $\sigma^{SI}_{DM-p} \sim m_{DM}^{-2}$, hence the scaled SI cross section is approximately constant.

For $m_{\widetilde{\nu}_R} \gtrsim 100$ GeV the annihilation mechanism involving pseudo-scalar particles becomes inefficient due to its dependence with the sneutrino mass, resulting in overproduction of DM. However, for $m_{\widetilde{\nu}_R}\sim 130-190$ GeV, solutions involving coannihilation with the Higgsino dominated neutralino are obtained. In this case, despite having points with $\lambda_N$ in the same range as in the previous region, the suppression coming from a higher $m_{\widetilde{\nu}_R}$ results in a lower $\sigma^{SI}_{\widetilde{\nu}_R}$.

To exemplify this and other mechanisms found, several benchmark points with RH sneutrino as DM candidate are shown in Appendix~\ref{sec:benchmarkpoints}. 
In Table~\ref{BPcoannihilation1} \textbf{BP1-4}  consider coannihilations with Higgsino, Wino, Singlino, and Bino dominated neutralinos, respectively. In Table~\ref{BPcoannihilation2} we show an example for coannihilation with stop in \textbf{BP5}. Finally, \textbf{BP6-8}, shown in Table~\ref{BPresonances}, we present the parameters for a RH sneutrino that annihilates via direct quartic coupling to a pair of light pseudo-scalars, a resonance with the SM-like Higgs boson, and a resonance with the second lightest Higgs boson, $h_2$, correspondingly.

Regarding the viability of the DM candidates of the model, we can conclude the following: 
\begin{itemize}
\item As expected, for neutralinos with masses below $\sim 100$ GeV, resonant conditions through $Z$ and $h_i$ are dominant, especially for Bino and Singlino-like neutralinos. They are depicted as vertical narrow strips. A significant Higgsino fraction is usually needed, although if the $h_2$ Higgs boson has dominant singlet component, some pure Singlinos can be found via funnel using the $\kappa \, \hat{S}^3$ term (see for example, benchmark point \textbf{BP3} in Table~\ref{BPcoannihilation1}). 
\item For Bino and Singlino neutralinos in the upper middle mass region, annihilations involving $A_i$ are also present. In the same region, Bino-like neutralino coannihilation channels with stops are relevant. 
\item As expected, Wino and Higgsino dominated neutralinos result in a low relic density for masses below $\sim 1$ TeV. They are arranged in two easily identifiable curves, and points with a mixture of Wino and Higgsino-like neutralinos lie between both sets. Below $\sim 100$ GeV these kind of neutralinos cannot be found due to chargino constraints.
\item RH sneutrinos with very low relic density 
can be found in the entire mass range, particularly through coannihilations and resonances.
\item For $m_{\widetilde{\nu}_R} \lsim 100$ GeV annihilations via direct quartic couplings to a pair of light CP-odd Higgs, and resonances to CP-even Higgs are the dominant channels to obtain an allowed RH sneutrino DM relic abundance. The former mechanism is possible because we are considering an NMSSM-like model.

\item For $m_{\widetilde{\nu}_R} \gsim 100$ GeV annihilations via direct quartic couplings are still significant, but coannihilations with Higgsino neutralino are dominant. Nonetheless, coannihilation with all kind of neutralinos (especially with Wino neutralinos) can be found. Another important contribution comes from resonances with the second lightest CP-even scalar, especially for low relic densities where it can be the only channel available.

\item For high RH sneutrino masses, coannihilation with sbottoms and stops are possible. For $m_{\widetilde{\nu}_R} \gsim 1200$ GeV coannihilations with the mentioned sparticles are dominant. We would like to mention that coannihilations with staus and gluinos are feasible for RH sneutrino masses \gsim 430 GeV and 2 TeV, respectively~\cite{ATLASllcharged:2019}, but are not present due to our choise of parameter values. 
\end{itemize}

\subsection{Direct and Indirect detection constraints}
\label{sec:DDandID}

\begin{figure}[t!]
\begin{center}
 \begin{tabular}{cc}
 \hspace*{-14mm}
 \epsfig{file=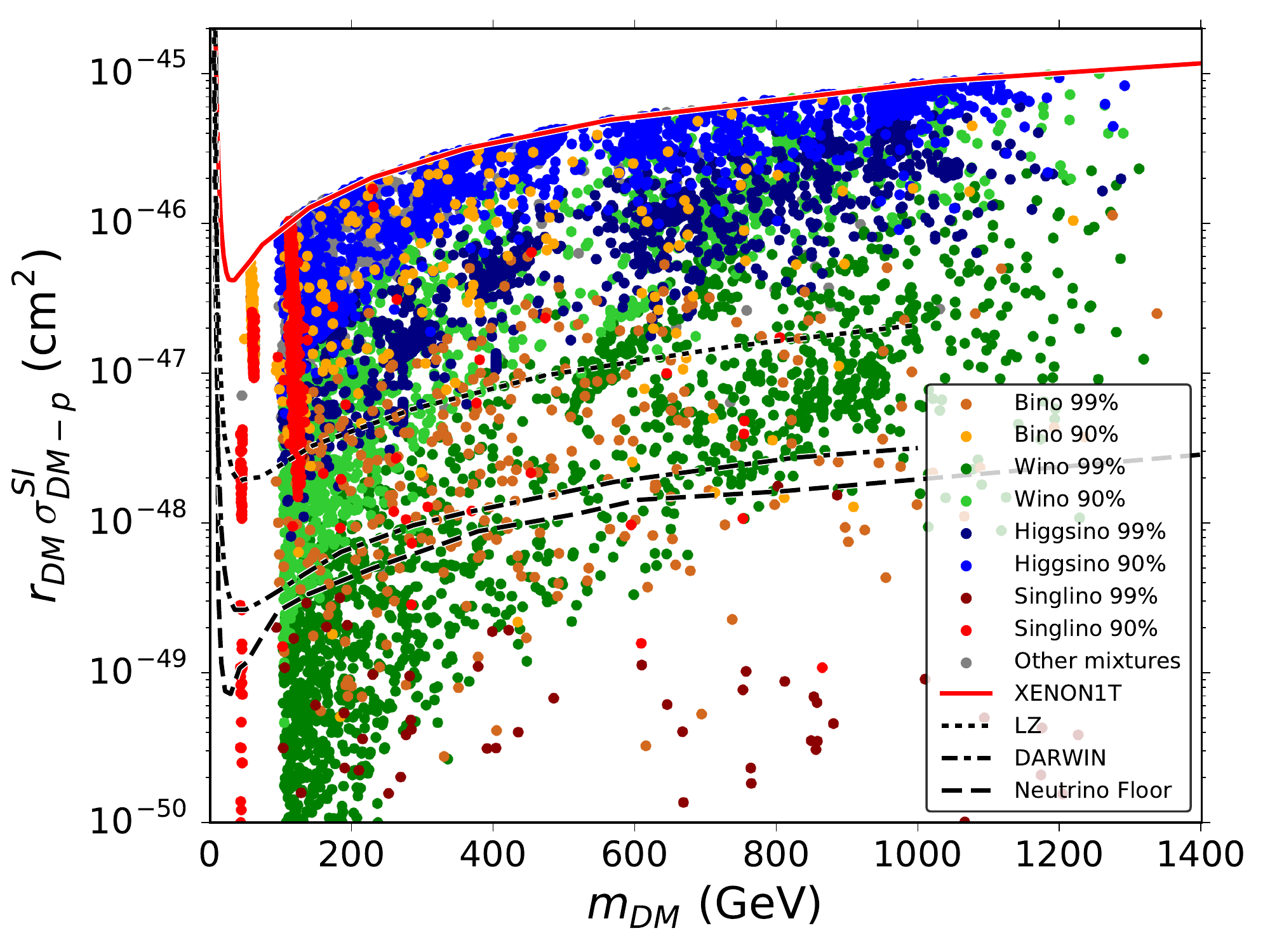,height=7cm} 
        \hspace*{-3mm}\epsfig{file=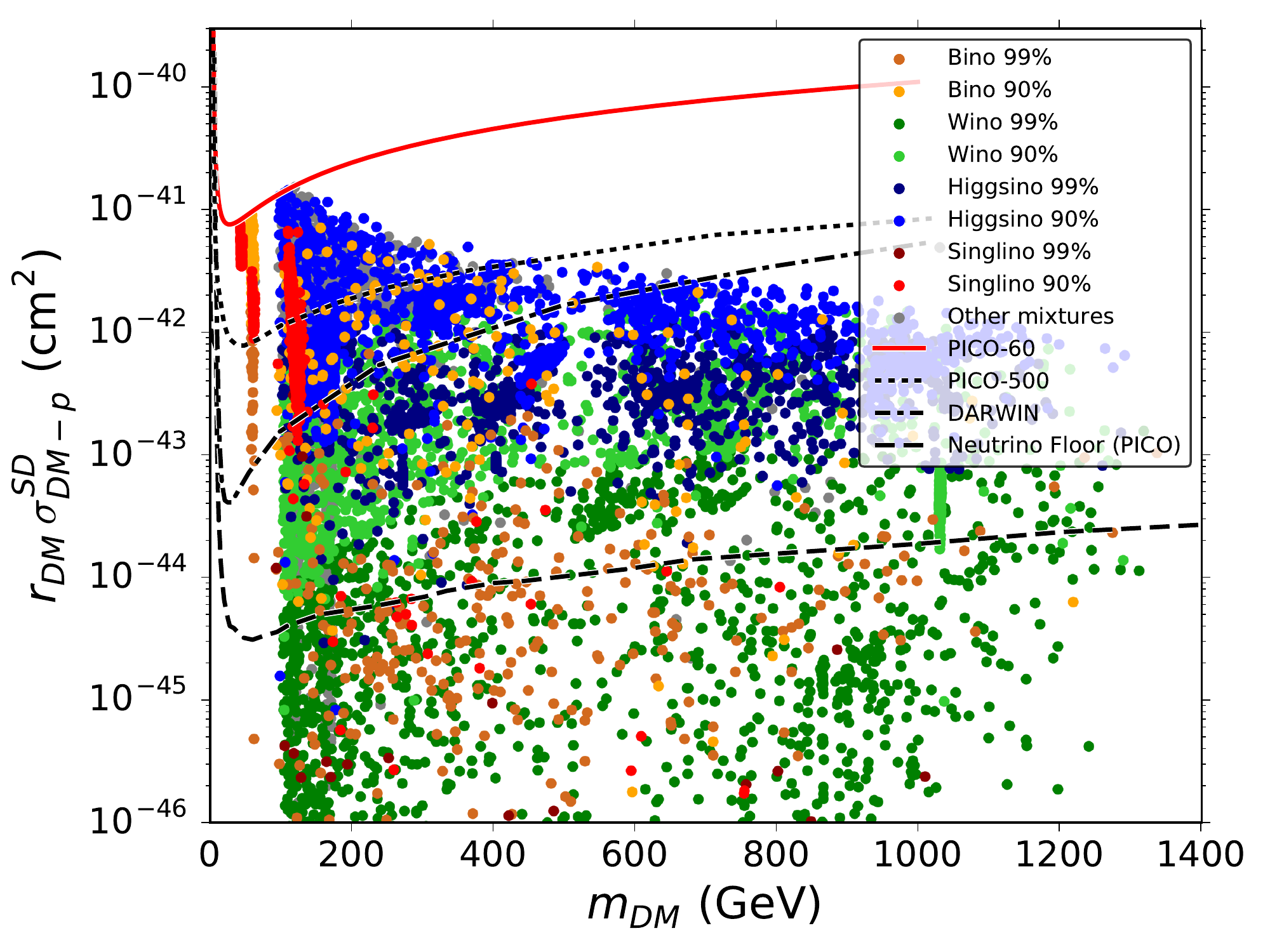,height=7cm} \\
 \hspace*{-14mm} \epsfig{file=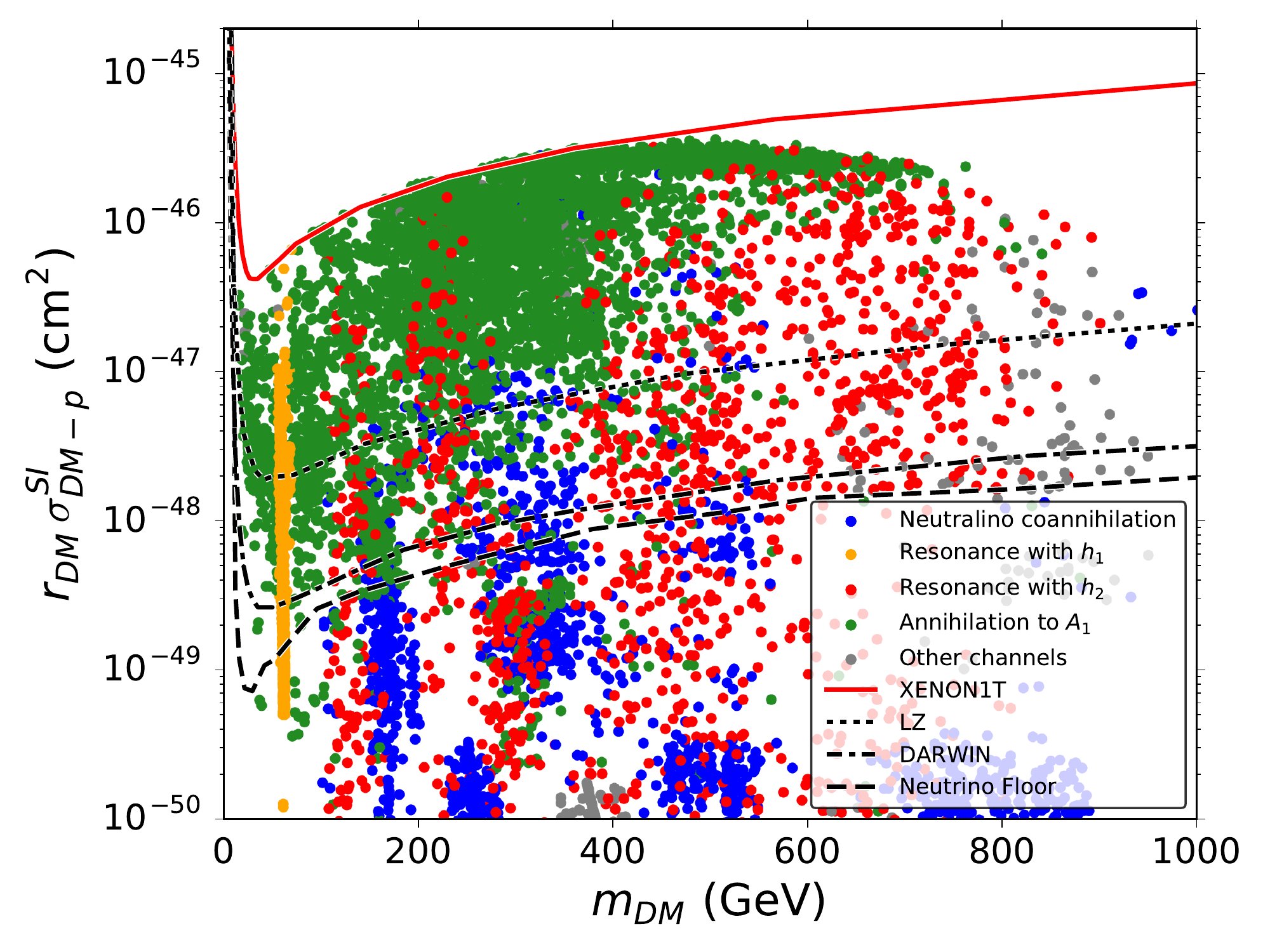,height=7cm} 
    \end{tabular}
    \captions{The first row presents the scaled spin-independent (top-left) and spin-dependent (top-right) direct detection cross sections for neutralino DM, the color coding represents its composition. The bottom row shows the scaled spin-independent direct detection cross sections for RH sneutrino DM; the color coding corresponds to the main channels used by RH sneutrinos to obtain an allowed abundance. The solid red curves show current experimental sensitivities from XENON1T~\cite{xenon1tcite1,xenon1tcite2} and PICO-60~\cite{picocite1,picocite2} for SI and SD, respectively. Projected sensitivities for DARWIN~\cite{DARWIN:2016}, LZ~\cite{LZ:2015} and PICO-500~\cite{PICO-500:2017} experiments are shown as black dotted and dot-dashed curves. The black dashed curves show the neutrino floor taken from Ref.~\cite{Billard:2014,Ruppin:2014}; for the SD case, the neutrino background depicted corresponds to experiments using $C_3F_8$ as detector material, like PICO. The scaling factor $r_{DM}$ accounts for the possibility that the calculated thermal relic density lies below the Planck measurement.
}
    \label{SIandSDscan}
\end{center}
\end{figure}

In this section, we present the impact of current and near future direct and indirect detection DM experiments on the model. In Fig.~\ref{SIandSDscan} we show the predicted scaled spin-independent (SI) and scaled spin-dependent (SD) scattering cross sections of the LSP with a target nucleus for the allowed points of our scan. The first row of figures corresponds to neutralino DM, and the second row to RH sneutrino DM. The scaling factor $r_{DM}=\frac{\Omega_{DM}h^2}{\Omega_{cdm}^{\text{Planck}}h^2}$ allow us to compare solutions with relic density below the Planck measurement against published constraint. This is particularly relevant for low mass Higgsinos and Winos. We also present the current experimental constraints from XENON1T~\cite{xenon1tcite1,xenon1tcite2} and PICO-60~\cite{picocite1,picocite2}, and the neutrino background floor~\cite{Billard:2014,Ruppin:2014} for SI and SD cross sections.

The projected sensitivities from DARWIN~\cite{DARWIN:2016}, LZ~\cite{LZ:2015} and PICO-500~\cite{PICO-500:2017} are also shown in Fig.~\ref{SIandSDscan}. Upcoming experiments will be able to probe an important region of the neutralino DM parameter space, for example, almost all the predicted points with Higgsinos could be tested by LZ. On the other hand, DARWIN will be important to probe the available parameter space up to the neutrino floor for SI, covering also the majority of Winos, Binos and Singlinos predicted. The SD neutrino background depends heavily on the material of the detector, (see Ref.~\cite{Ruppin:2014}). A significant region will be tested by PICO-500 and DARWIN. As in the SI case, this is especially true for Higgsinos, as well as for Bino and Singlino resonances with Z and light $h_i$.

On the other hand, next generation experiments will also probe RH sneutrino DM. In the bottom panel of Fig.~\ref{SIandSDscan} the same color coding as in the left panel of Fig.~\ref{relicscanRHsneutrinos} is used, representing the main channels used by RH sneutrinos to obtain an allowed abundance. As can be seen, almost all points that would be explored by LZ and DARWIN do not coannihilate with neutralinos. In fact, the dominant mechanism in this region involves annihilations to a pair of pseudo-scalars through direct quartic coupling. For $m_{\tilde{\nu}_R} \gtrsim 500$ GeV resonances with heavy CP-even Higgs scalar are important. 

\begin{figure}[t!]
\begin{center}
 \begin{tabular}{cc}
 \hspace*{-14mm}
 \epsfig{file=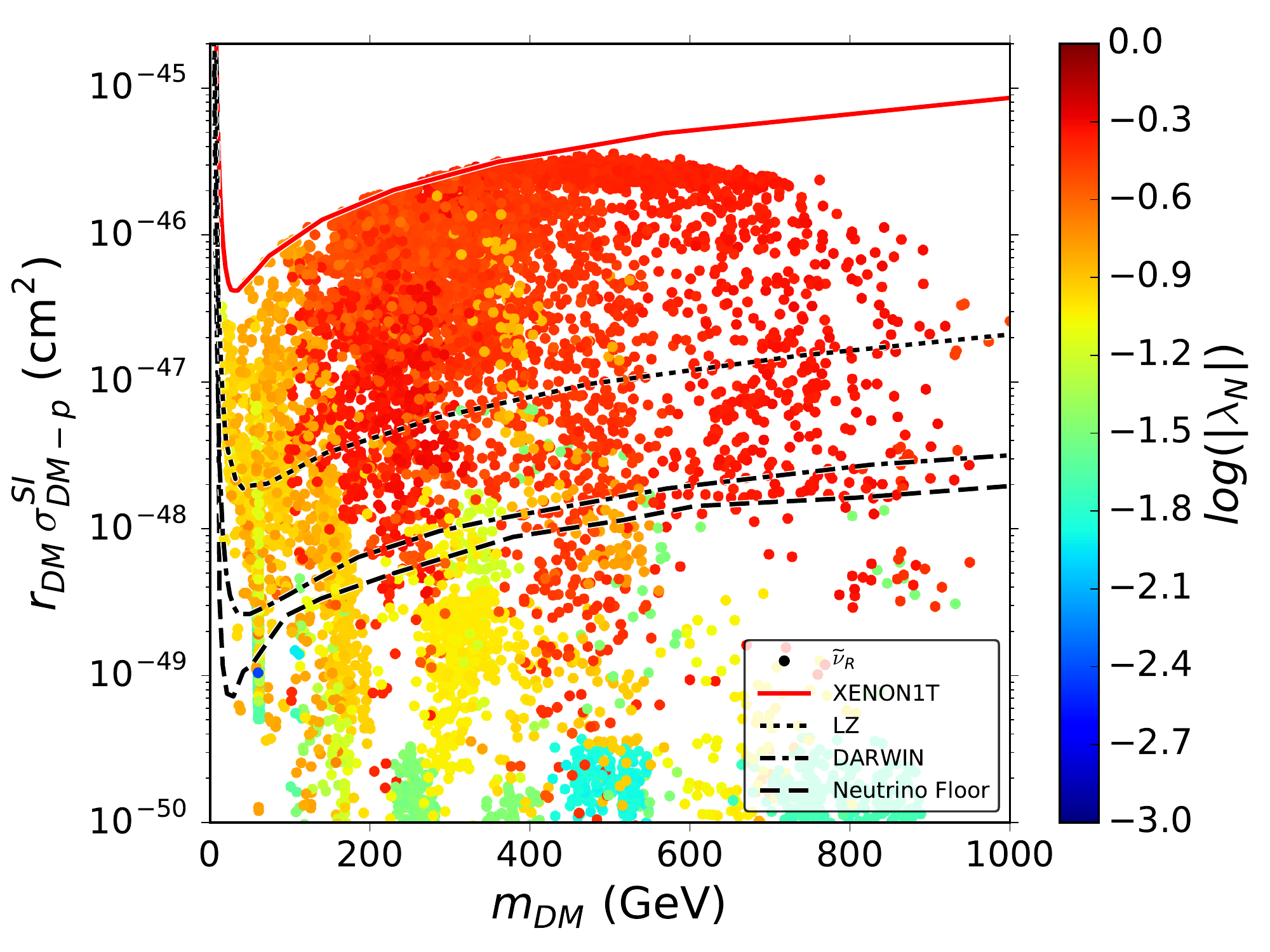,height=7cm} 
        \hspace*{-2mm}\epsfig{file=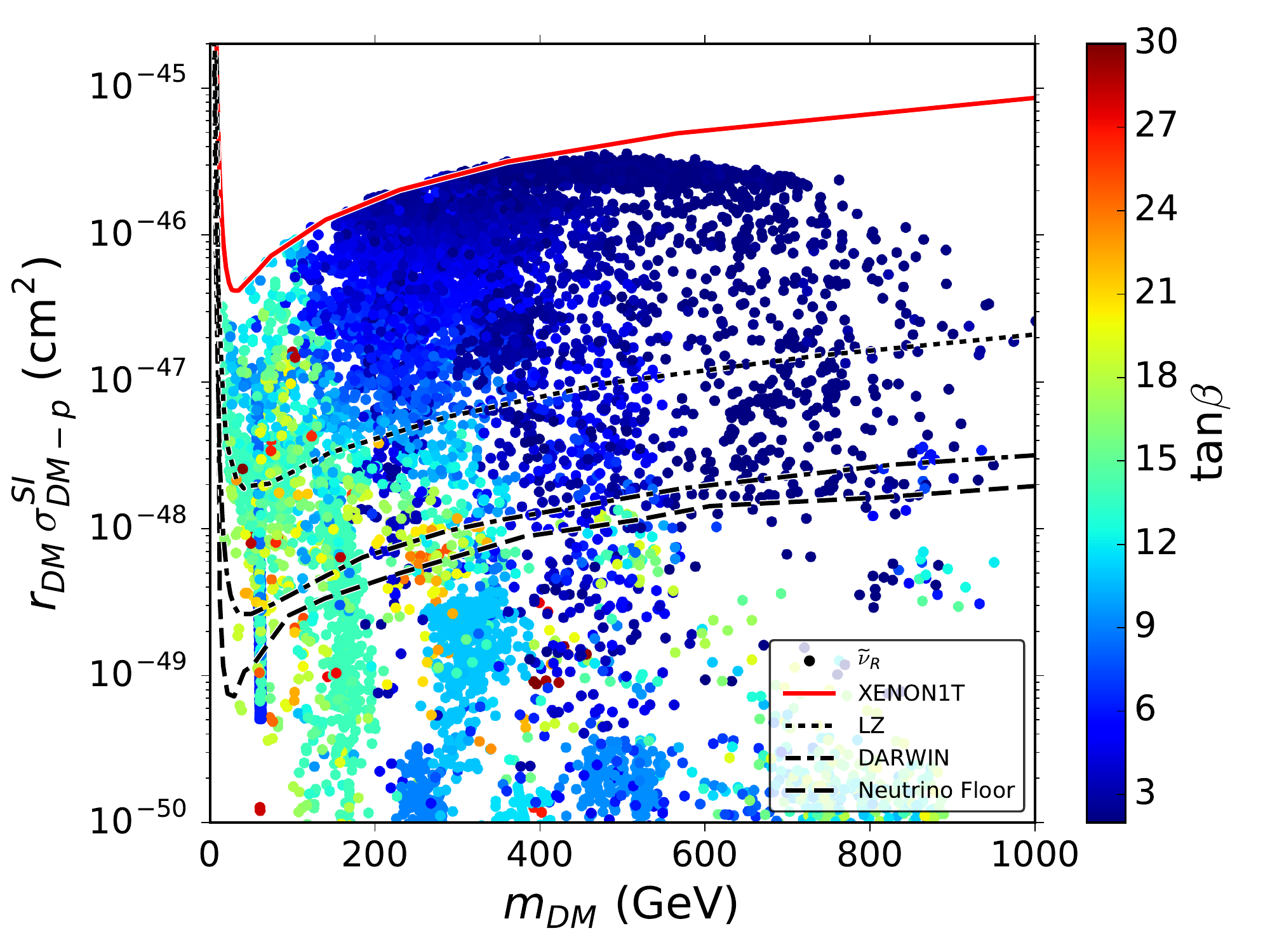,height=7cm}   
    \end{tabular}
    \captions{Scaled spin-independent direct detection cross sections for RH sneutrino DM; the color coding corresponds to the value of $\lambda_N$ (left), and the value of $tan \, \beta$ (right) for each point. The rest of the references are the same as in Fig.~\ref{SIandSDscan}.
    }
    \label{SIandSDscan2}
\end{center}
\end{figure}

Solutions with RH sneutrino DM using coannihilations with neutralinos prefer low values of $|\lambda_N^3|$ and $\lambda$, therefore tend to produce very small SI cross section values (see the SI cross section approximation shown in Eq.~(\ref{sigmaSIapprox1})). In our scan, we allow values as low as 0.001 for these inputs, resulting in very small $\sigma^{SI}_{\widetilde{\nu}_R-p}$. For increasing DM masses, most solutions with coannihilations tend to cut deeper into the neutrino coherent scattering background, making them very challenging to test with current techniques. Nonetheless, some points that coannihilate with neutralinos lie in the region to be probed by LZ and DARWIN. In general, these solutions also present contributions from annihilations to a pair of pseudo-scalars or resonances with $h_i$. These mechanisms allow larger values of $|\lambda_N^3|$, as can be seen in the left panel of Fig.~\ref{SIandSDscan2}.

The other two relevant parameters also involved in the RH sneutrino SI cross section estimate are $m_{DM}$ and $\tan \beta$. On the right panel of Fig.~\ref{SIandSDscan2} the color coding shows the value of $\tan \beta$, and as expected from Eq.~(\ref{sigmaSIapprox1}), we get decreasing values of $\tan \beta$ for increasing values of $m_{DM}$. The impact of both parameters is counteracted to keep $ r_{DM} \times \sigma^{SI}_{\widetilde{\nu}_R-p} \sim $ constant. Finally, we would like to remark that the region with RH sneutrino DM that could be explored by next generation experiments prefers low values of $\tan \beta$, in particular for $m_{\tilde{\nu}_R} \gtrsim 200$ GeV we get  $\tan \beta \lsim 10$. In that regard, for $m_{\tilde{\nu}_R} \gtrsim 500$ GeV, it is dificult to obtain a solution with $\sigma^{SI}_{\widetilde{\nu}_R-p}$ up to the XENON1T bound, because $\lambda$ and $\lambda_N$ ($\tan \beta$) need to take values close to their allowed upper (lower) limits (see Table~\ref{scanparameters} and Eq.~(\ref{sigmaSIapprox1})).

Although SI experiments provide a more sensitive tool according to the projected experiments, a joint SI and SD analysis offers the opportunity to disentangle the LSP identity between the two DM candidates of the model. 
For example, for coannihilation points, a slight variation of the free parameters in the sneutrino sector can give us RH sneutrino DM or neutralino DM with approximately the same mass. This situation could be unraveled as coannihilating RH sneutrinos tend to have a very small SI cross section. On the other hand, if we consider RH sneutrino DM that annihilates through a quartic coupling, we can get a similar SI signature to the case of neutrino DM, but with vanishing SD signal. Of course, in the case of neutralino DM, a combination of SI and SD experiments can help determine its composition.

\begin{figure}[t!]
\begin{center}

 \epsfig{file=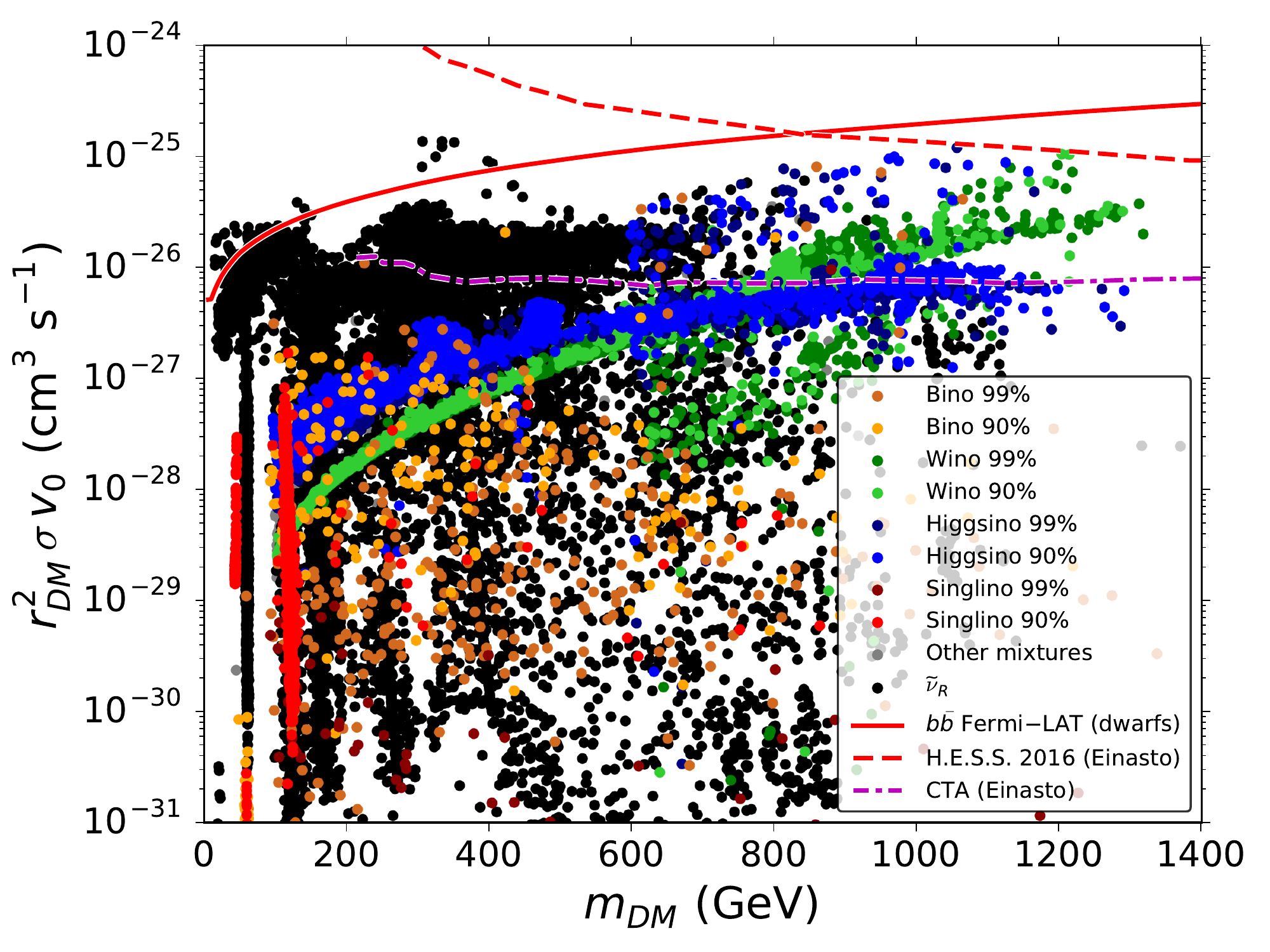,height=9cm} 

    \captions{Distribution of the scanned points in the ($m_{DM}$, $r_{DM}^2 \, \langle \sigma v_0\rangle$) space, where $r_{DM}$ is the DM relic density fraction. The color coding represents the LSP identity, and the dominant composition in the case of neutralino, as in Fig.~\ref{relicscan}. The current upper 95\% C.L. limits from \Fermi-LAT to $b \bar{b}$ from the observation of dwarfs~\cite{Ackermann:2015dwarfs}, and the limits of H.E.S.S. from observations of the Galactic center using Einasto profile~\cite{Hryczuk:2019} are indicated as solid and dashed red curves, respectively. The projected CTA sensitivity~\cite{Hryczuk:2019} is shown as a dot-dashed magenta curve.
}
    \label{IDscan}
\end{center}
\end{figure}

To analyse the impact of current and future indirect detection experiments, in Fig.~\ref{IDscan}, we show the normalized thermally averaged annihilation cross section, $r_{DM}^2 \times \langle \sigma v_0\rangle$, as a function of the DM mass, where $r_{DM}$ is the DM relic density fraction defined previously.

Upper bounds on annihilation cross sections are derived for pure channels. In our scan we obtain a mixture of several annihilation channels, therefore the data sets were implemented on each of the reported channels. In case of annihilation final states for which limits have not been reported by the collaborations, we employ the most relevant existing bounds. In particular, for the $e^-e^+$ channel we consider the same limit as for $\mu^-\mu^+$, for $u \bar{u}$, $d \bar{d}$, $s \bar{s}$, $c \bar{c}$, $A_1 A_1$, $Z Z$ and $h h$ we use $b \bar{b}$, and for $H^-H^+$ the bound for $W^+W^-$.

It is worth noticing that in our analysis we apply indirect detection constraints at face value because direct detection limits are usually more restrictive to set the allowed parameter points. However uncertainties associated with DM density profiles and astrophysical background modeling result in bounds on the annihilation cross section that can vary up to an order of magnitude~\cite{Silverwood:2014yza,Bergeron:2017rdm}.

In Fig.~\ref{IDscan}, the solid red curve corresponds to the limit of $b \bar{b}$ final state set by \Fermi-LAT from the observation of dwarfs galaxies~\cite{Ackermann:2015dwarfs}, to guide the eye. The dashed red curve obtained by H.E.S.S. collaboration~\cite{HESS:2016} represents annihilation lower limits from observations of the Galactic center using Einasto profile calculated in Ref.~\cite{Hryczuk:2019}. The dot-dashed magenta curve corresponds to the projected CTA~\cite{CTA:2017} 95\% C.L. sensitivity to DM annihilation derived from observations of the Galactic center assuming 500 hour homogeneous exposure and Einasto profile~\cite{Hryczuk:2019}. As can be seen, a small fraction of points will be explored by CTA. In the case of neutralino DM, the region that would be probed is constituted of mostly Wino dominated points with $r_{DM}>0.1$, i.e. with masses $\gsim 750$ GeV. Considering RH sneutrino DM, CTA will study points with dominant annihilation channel involving direct quartic coupling to pseudo-scalars, and masses up to $\sim 800$ GeV.

\begin{figure}[t!]
\begin{center}

        \hspace*{-6mm}\epsfig{file=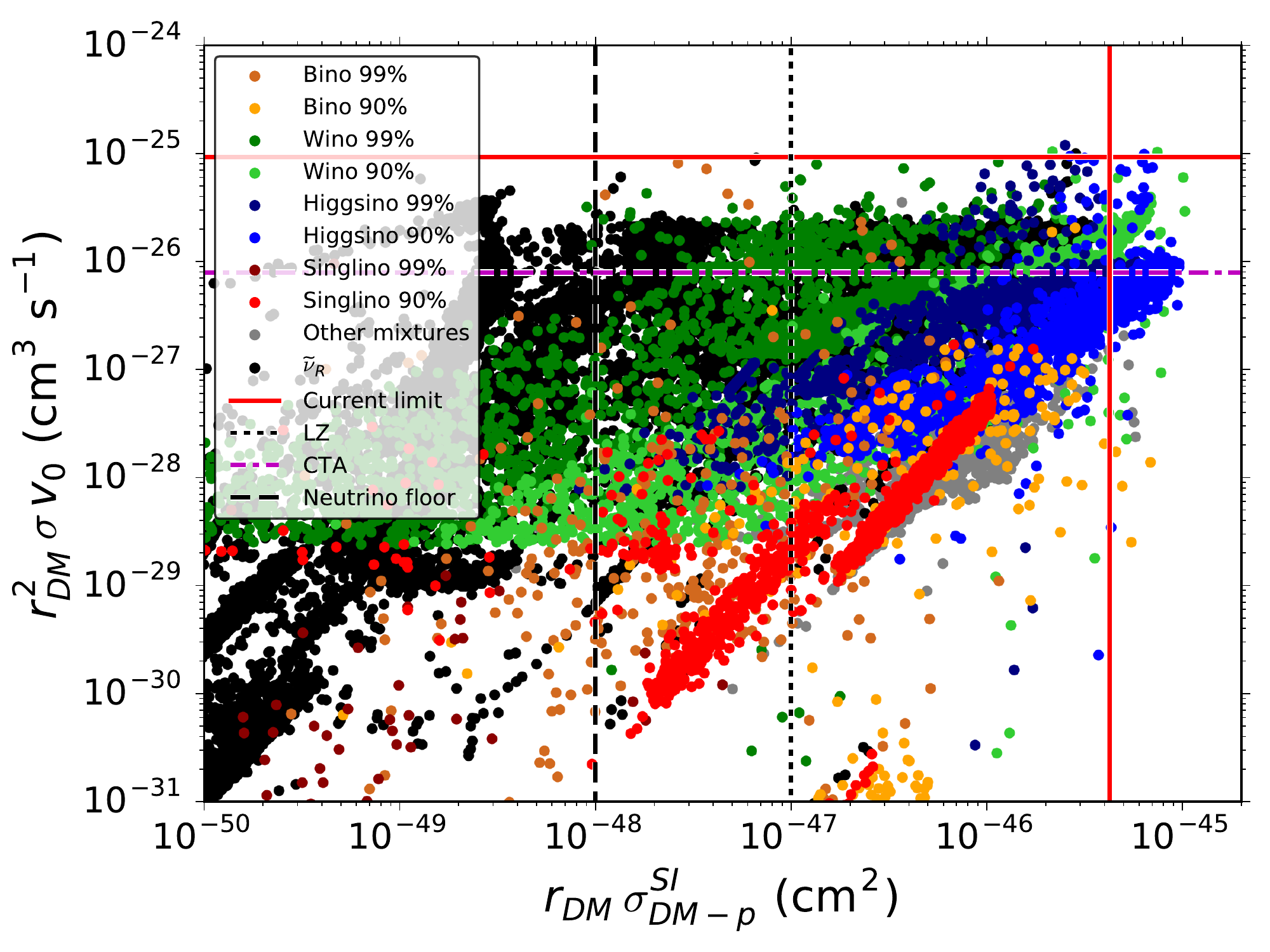,height=9cm}   

    \captions{Direct and indirect detection complementarity. Scanned points in the ($r_{DM} \, \sigma_{DM-p}^{SI}$, $r_{DM}^2 \, \langle \sigma v_0\rangle$) plane with the same color coding as in Fig.~\ref{relicscan}. XENON1T and \Fermi-LAT set the current direct and indirect detection upper limits, respectively. As representative values for the neutrino floor, current upper limits and projected sensitivity we consider $m_{DM}\approx 500$ GeV.
}
    \label{DDandIDscan}
\end{center}
\end{figure}

Finally in Fig.~\ref{DDandIDscan} we present the complementarity of direct and indirect detection experiments and the future prospects of detection. The experimental constraints consider $m_{DM}\approx 500$ GeV to help the reader. While the next generation of direct detection experiments will cut deep into the parameter space, a complementary approach would be important to determine the characteristics of dark matter, especially for Winos and RH sneutrinos.


\section{Conclusions}
\label{sec:conclusions}

In this work we have analysed the next-to-minimal supersymmetric standard model including right-handed neutrinos superfields.
In this framework, besides the usual neutralino, the RH scalar partner of the neutrino becomes a good DM candidate.

As RH sneutrinos do not interact directly with SM particles, we would typically expect DM overproduction and very small SI scattering cross section with nucleons, even below the neutrino background. However, one crucial feature of the model is the existence of RH sneutrino direct couplings to Higgs bosons and to neutralinos. This fact arises due to interaction terms between RH neutrino and singlet fields ($\lambda_N S N N$, and the corresponding soft-breaking terms), absent in the NMSSM. Then, the extra terms are key factors to obtain the correct amount of relic density, and to allow small but measurable DM-nucleon scattering and DM annihilation cross section.

To explore the model, we have carried out a scan of the parameter space employing \texttt{Multinest} as optimizer. Several constraints have been imposed, including the measured amount of DM in the Universe, direct and indirect DM detection experiments, Higgs data, flavor physics, and SUSY searches at colliders probing light stops, sbottoms, charginos and leptons plus missing $E_T$ in the final state. Additionally, we have analysed the impact and complementarity of near future direct and indirect detection endeavours.

Although in our scan we have used a likelihood data-driven method, we have not performed an statistical interpretation as done in Ref.\cite{Cao:2019}. Instead, we have identified regions with viable DM candidates and we have studied their characteristic features for a broader parameter space, which in turn results in a wider DM mass range allowed. Moreover, we have not assumed that the thermal relic density saturates the experimental value, allowing multicomponent DM scenarios in order to be as general as possible. 

We have studied both neutralino and RH sneutrino as DM candidates, and concluded that viable solutions can be obtained for a wide range of DM masses, from a few GeV to above 1 TeV, over a broad parameter space. Since the neutralino behaves similarly to the NMSSM one, we have emphasized our analysis on the RH sneutrino. We have found that to get an allowed relic abundance, RH sneutrino mainly uses three mechanisms to be in equilibrium with the thermal bath in the early Universe: resonances with CP-even Higgs, annihilations through scalar quartic couplings with two CP-odd Higgs in the final state, and coannihilations with Higgsinos.

Resonances with CP-even Higgs bosons are well-known features, and one of the main channels used for low mass RH sneutrinos, $m_{\tilde{\nu}_R}<100$ GeV. However, they are still important for heavier masses especially when $h_2$ is singlet dominated.

Annihilations through direct quartic couplings to a pair of pseudo-scalars, whose importance has been overlooked in previous studies, play a very important role for the entire mass range. Solutions with these mechanisms, can satisfy in a natural way current constraints and lie in a region that would be tested by next generation direct and indirect detection experiments, providing a very appealing prospect. For $m_{\tilde{\nu}_R}<100$ GeV, these annihilation channels are kinematically allowed as CP-odd Higgs states with significant singlet component can be very light in NMSSM-like models.

Finally, most solutions present coannihilations with Higgsinos. Furthermore, coannihilations with all kind of neutralino have been found, even with another sparticles like stops. It is worth mentioning that significant contributions come from coannihilation with Winos, also overlooked in previous works. These mechanisms require some fine tuning between the RH sneutrino and its coannihilation partner, i.e. a mass difference $\lesssim 10 \%$. Then, we have seen that the former particle inherits restrictions and properties from the latter, in particular, the RH sneutrino relic abundance lower limit is determined by the relic density of the corresponding neutralino. 

Another crucial feature of the model is that the parameters involved in the RH sneutrino mass, like $m_{\tilde{N}}$, $\lambda_N$, and $T_{\lambda_N}$, do not affect the rest of the mass spectrum. In this way, we can find easily parameter points with RH sneutrinos DM through coannihilations, using viable neutralino points as seeds. Nonetheless, the mentioned parameters are involved in the DM-nucleon scattering rate, which generally results in extremelly low SI cross sections beyond the sensitivity of next generation experiments for heavy RH sneutrinos, unless significant contributions of other annihilation mechanisms are also present, like the mentioned resonances or quartic coupling channels.

\pagebreak


\section*{Acknowledgments}

The work of DL and AP was supported by the Argentinian CONICET, and also acknowledges the support through PIP11220170100154CO. They would like to thank the team supporting the Dirac High Performance Cluster at the Physics Department, FCEyN, UBA, for the computing time and their dedication in maintaining the cluster facilities. DL also acknowledges the Spanish grant PGC2018-095161-B-I00.
R. RdA acknowledges partial funding/support from the Elusives ITN (Marie Sklodowska-Curie grant agreement No 674896), the ``SOM Sabor y origen de la Materia" (FPA 2017-85985-P). The authors thank C. Muñoz for useful comments.


\appendix

\section{Benchmark points: RH sneutrino as DM candidate}
\label{sec:benchmarkpoints}

\begin{table}
\begin{footnotesize}
\begin{center}
\begin{tabular}{|c|c|c|c|c|}

        \hline
        \textbf{Parameters} & \textbf{BP1} & \textbf{BP2} & \textbf{BP3} & \textbf{BP4} \\
        \hline
        \hline
            $M_1$ & 2388.7 GeV & 1955.4 GeV & 1229.5 GeV & 379.1 GeV \\
            $M_2$ & 2837.9 GeV & 546.4 GeV & 294.5 GeV & 2190.2 GeV \\
            $\mu_{eff}$ & 267.6 GeV & 1273.9 GeV & 159.8 GeV & 474.8 GeV \\
            $\tan \beta$ & 9.009 & 12.32 & 11.09 & 10.07 \\
            $\lambda$ & 0.0646 & 0.0220 & 0.0717 & 0.00641 \\
            $\kappa$ & 0.138 & 0.0120 & 0.0283 & 0.0382 \\
            $\lambda_N^3$ & -0.0327 & -0.00869 & -0.00560 & -0.00181 \\
            $T_{\lambda}$ & 0.121 GeV & 0.288 GeV & 134.8 GeV & 2.00 GeV\\
            $T_{\kappa}$ & -3.10 GeV & -0.0742 GeV & -0.105 GeV & -1.42 GeV \\
            $T_{\lambda_N}^3$ & -0.258 GeV & -0.114 GeV & -3.55 GeV & -3.26 GeV \\
            $m^2_{\tilde{N}_3}$ & 1.53$\times 10^{5}$ GeV$^2$ & 3.27$\times 10^{4}$ GeV$^2$ & 3.10$\times 10^{4}$ GeV$^2$ & 1.34$\times 10^{6}$ GeV$^2$ \\
            $m^2_{\tilde{u}_3}$ & 3.31$\times 10^{6}$ GeV$^2$ & 1.57$\times 10^{6}$ GeV$^2$ & 1.76$\times 10^{6}$ GeV$^2$ & 2.46$\times 10^{6}$ GeV$^2$ \\
            $m^2_{\tilde{Q}_3}$ & 1.64$\times 10^{6}$ GeV$^2$ & 3.80$\times 10^{6}$ GeV$^2$ & 1.39$\times 10^{6}$ GeV$^2$ & 3.20$\times 10^{5}$ GeV$^2$ \\
            $T_{u_3}$ & 2223.3 GeV & 2475 GeV & 2239 GeV & 2144.5 GeV \\
        \hline
        \hline
        \textbf{Spectrum} & & & & \\
        \hline
        \hline
            $m_{\tilde{\nu}_{R}}$ & 259.8 GeV & 546.0 GeV & 119.3 GeV & 366.8 GeV \\
            $m_{\chi_1^0}$ & 272.1 GeV & 566.6 GeV & 124.0 GeV & 366.9 GeV\\
            $N_{11}^2$ & 2.7$\times 10^{-4}$ & 1.9$\times 10^{-6}$ & 1.6$\times 10^{-4}$ & 0.91 \\
            $N_{12}^2$ & 5.6$\times 10^{-4}$ & 0.99 & 0.017 & 7.7$\times 10^{-5}$ \\
            $N_{13}^2+N_{14}^2$ & 0.99 & 7.3$\times 10^{-3}$ & 0.16 & 0.088 \\
            $N_{15}^2$ & 6.3$\times 10^{-5}$ & 1.0$\times 10^{-7}$ & 0.82 & 3$\times 10^{-9}$ \\
            $m_{\chi_2^0}$ & 274.7 GeV & 1280.4 GeV & 145.2 GeV & 479.9 GeV \\
            $m_{\chi_1^{\pm}}$ & 273.6 GeV & 566.8 GeV & 146.9 GeV & 478.2 GeV \\
            $m_{h_1}$ & 122.7 GeV & 126.4 GeV & 124.9 GeV & 125.0 GeV \\
            $m_{h_2}$ & 1183.8 GeV & 1489.7 GeV & 252.6 GeV & 3965.4 GeV \\
            $S_{23}^2$ & 0.95 & 0.99 & 0.99 & 1.4$\times 10^{-7}$ \\
            $m_{A_1}$ & 197.2 GeV & 115.6 GeV & 28.1 GeV & 573.5 GeV \\
            $P_{12}^2$ & 0.99 & 0.99 & 0.99 & 0.99\\
        \hline
        \hline
        \textbf{Dark Matter} & $\tilde{\nu}_{R}$ & $\tilde{\nu}_{R}$ & $\tilde{\nu}_{R}$ & $\tilde{\nu}_{R}$ \\
        \hline
        \hline
            $\Omega_{\text{DM}}h^2$ & 0.0409 & 0.0431 & 0.0146 & 0.121\\
            $\sigma_{DM-p}^{SI}$ & 2.6$\times 10^{-50}$ cm$^2$ & 7.4$\times 10^{-53}$ cm$^2$ & 9.7$\times 10^{-51}$ cm$^2$ & 5.2$\times 10^{-56}$ cm$^2$\\
        
            main & $\chi_{1,2}^0 \, \chi_{1,2}^0 \rightarrow$ SM & $\chi_1^0 \, \chi_1^0 \rightarrow$ SM & $\tilde{\nu}_{R} \, \tilde{\nu}_{R} \rightarrow A_1 \, A_1$ $(\sim 52\%)$ & $\tilde{q} \, \tilde{q} \rightarrow$ SM $(\sim 65\%)$ \\
            annihilation & $\chi_{1,2}^0 \, \chi_1^{\pm} \rightarrow$ SM & $\chi_{1}^0 \, \chi_1^{\pm} \rightarrow$ SM & $\tilde{\nu}_{R} \, \tilde{\nu}_{R} \rightarrow$ SM $(\sim 34\%)$ & $\tilde{q} \, \chi_{1}^0 \rightarrow$ SM $(\sim 25\%)$ \\
            channels & $\chi_1^{\pm} \, \chi_1^{\pm} \rightarrow$ SM & $\chi_1^{\pm} \, \chi_1^{\pm} \rightarrow$ SM & $\chi_1^0 \, \chi_1^0 \rightarrow$ SM $(\sim 11\%)$ & $\chi_{1}^0 \, \chi_{1}^0 \rightarrow$ SM $(\sim 8\%)$ \\
        \hline
\end{tabular}
\end{center}
\end{footnotesize}
  \caption{Set of inputs for several benchmark points with RH sneutrino DM through coannihilation mechanism with neutrinos. $N$, $S$ and $P$ are the mass mixing matrixes of the neutralino, the CP-even Higgs and CP-odd Higgs sectors, respectively.}
  \label{BPcoannihilation1}
\end{table}

In this appendix, we show some benchmark points to illustrate different annihilation channels that RH sneutrinos can have to get an allowed relic density. 

We begin showing coannihilations with neutralinos. This mechanism depends heavily on the type of coannihilation partner, which has to annihilate efficiently, since RH sneutrino relic abundance have the would-be relic density of the former as lower limit. Higgsinos or Winos LSP with $m_{\chi_1^0}\lesssim 1$ TeV usually result in low relic abundances, and at the same time, can evade the stringent direct detection constraints. On the other side, Binos can have relic abundance within current constraints through resonances, or coannihilations with heavy squarks. Finally, unless a resonance is present, Singlinos tend to have huge relic abundances due to its small coupling with the rest of the particles.

For those reasons, and the presence of the terms $\lambda_N S N N$ and $\lambda S H_u H_d$ in the superpotential, coannihilation with Higgsino via s-channel singlet exchange is the main process to get viable RH sneutrino DM. Coannihilations with Winos are also relevant and a significant fraction of solutions use this mechanism. Nevertheless, to consider RH sneutrinos via Bino or Singlino coannihilation processes can be important to achieve the correct relic abundance, especially for low DM mass, where Higgsino and Wino neutralinos are constraint by chargino searches.

In Table~\ref{BPcoannihilation1} we show sets of inputs for several benchmark points with RH sneutrino DM through different coannihilation mechanisms with neutralinos. We would like to comment that in all the examples shown in this appendix $h_1$ is a SM-like Higgs.

\textbf{BP1:} Higgsino dominated neutralino coannihilation. With this benchmark point we show the dominant mechanism to generate RH sneutrino with an allowed amount of relic density.  RH sneutrino annihilates through neutralino-chargino coannihilation, both with a dominant content of Higgsino. As the Higgsino neutralinos and the Higgsino chargino have approximately the same mass, 
LEP constraint on chargino searches impose $m_{\chi_1^0}\gtrsim 100$ GeV. Therefore, these kind of RH sneutrinos can be found for 100 GeV $\lesssim m_{\tilde{\nu}_R}\lesssim 1.2$ TeV. 

\textbf{BP2:} Wino dominated neutralino coannihilation. Unlike the previous case, only the lightest neutralino and lightest chargino are involved, both with important Wino content. As before, these RH sneutrinos can be found for 100 GeV $\lesssim m_{\tilde{\nu}_R}\lesssim 1.5$ TeV.

\textbf{BP3:} Singlino dominated neutralino coannihilation. Only the lightest neutralino is involved. However, as the singlino neutralino uses the same resonant annihilation channels that can be used by the RH sneutrino, the coannihilation mechanisms is not the dominant one. In this example, the RH sneutrino annihilates to a pair of lightest pseudo-scalar particles dominantly (52\%), but the contribution from coannihilation channels is not negligible (more than 11\%) due to a not negligible Higgsino component.

We can probe the same parameters as in BP3 but choosing a higher value of $m^2_{\tilde{N}_3} = 3.00\times 10^{5}$ GeV$^2$. This only affects the sneutrino masses, which turns out to be higher and not the LSP, $m_{\tilde{\nu}_{R}}=532.2$ GeV, but leaves the rest of the mass spectrum unchanged. Then, the DM candidate is a singlino dominated neutralino, with $\Omega_{\text{DM}}h^2=0.0124$, that annihilates mainly to a pair of lightest pseudo-scalar particles (53\%).

Notice that in both cases the singlino neutralino fulfills the resonant condition with the second lightest CP-even Higgs scalar. It has $m_{h_2}\simeq 252.6$ GeV and dominant singlet composition. Unlike BP1 and BP2, the term $\kappa \, S \, S \, S$ is also involved in the singlino case.

\textbf{BP4:} Bino dominated neutralino coannihilation. Binos LSP usually give large relic density values. Hence low Binos need to fulfill a resonant condition to achieve an allowed relic abundance and evade direct detection constraints. For higher mass range, they can also use a coannihilation channel which demands more particles with same mass. In this benchmark point besides a Bino with non negligible Higgsino contribution, a low mass stop is present, then $m_{\tilde{\nu}_R} \sim m_{\chi_1^0} \sim m_{\tilde{t}}$. Therefore, collider signatures on compressed scenarios are important. Since we have the following mass hierarchy $m_{\tilde{\nu}_R} < m_{\chi_1^0} < m_{\tilde{t}}$, with $m_{\tilde{t}}=390$ GeV, then the stop would decay to a quark and the Bino neutralino. The latter would subsequently decay into RH sneutrino and SM particles, but due to phase space suppression the signal would consist of missing transverse energy, i.e. the constraints turn out to be the same as simply considering neutralino LSP without a lighter RH sneutrino.

\begin{table}
\begin{footnotesize}
\begin{center}
\begin{tabular}{|c|c||c|c|}

        \hline
        \textbf{Parameters} & \multicolumn{3}{c|}{\textbf{BP5}} \\
        \hline
        \hline
            $M_1$ & 1638.7 GeV & \hspace{0.7cm} $T_{\lambda}$ \hspace{0.7cm} & 0.00856 GeV \\
            $M_2$ & 2986.1 GeV & $T_{\kappa}$ & -0.618 GeV \\
            $\mu_{eff}$ & 1508.6 GeV & $T_{\lambda_N}^3$ & -1.818 GeV \\
            $\tan \beta$ & 9.118 & $m^2_{\tilde{N}_3}$ & 1.91$\times 10^{6}$ GeV$^2$ \\
            $\lambda$ & 0.0975 & $m^2_{\tilde{u}_3}$ & 3.80$\times 10^{6}$ GeV$^2$ \\
            $\kappa$ & 0.248 & $m^2_{\tilde{Q}_3}$ & 2.00$\times 10^{6}$ GeV$^2$ \\
            $\lambda_N^3$ & -0.00124 & $T_{u_3}$ & 3844.5 GeV \\
        \hline
        \hline
        \textbf{Spectrum} & \multicolumn{3}{c|}{ } \\
        \hline
        \hline
            $m_{\tilde{\nu}_{R}}$ & 1305.9 GeV & $m_{h_1}$ & 124.2 GeV \\
            $m_{\tilde{t}}$ & 1343.1 GeV & $m_{h_2}$ & 7451.1 GeV \\
            $m_{\tilde{b}}$ & 1418.6 GeV & $S_{23}^2$ & 3.3$\times 10^{-4}$ \\
            $m_{\chi_1^0}$ & 1504.2 GeV & $m_{A_1}$ & 150.9 GeV \\
            $N_{13}^2+N_{14}^2$ & 0.92 & $P_{12}^2$ & 0.99 \\
            $m_{\chi_1^{\pm}}$ & 1514.4 GeV & $m_{A_2}$ & 7450.5 GeV \\
        \hline
        \hline
        \textbf{Dark Matter} & \multicolumn{3}{c|}{ $\tilde{\nu}_{R}$ } \\
        \hline
        \hline
            $\Omega_{\text{DM}}h^2$ & \multicolumn{3}{c|}{0.1159} \\
            $\sigma_{DM-p}^{SI}$ & \multicolumn{3}{c|}{3.8$\times 10^{-54}$ cm$^2$ } \\
            main & \multicolumn{3}{c|}{$\tilde{t} \, \tilde{t}^* \rightarrow$ SM $(\sim 77\%)$} \\
            annihilation & \multicolumn{3}{c|}{$\tilde{t} \, \tilde{b} \rightarrow$ SM $(\sim 15\%)$} \\
            channels & \multicolumn{3}{c|}{ } \\
        \hline
\end{tabular}
\end{center}
\end{footnotesize}
  \caption{Same as \ref{BPcoannihilation1} but benchmark point with RH sneutrino DM through coannihilation mechanism with stops.}
  \label{BPcoannihilation2}
\end{table}

In Table~\ref{BPcoannihilation2} we show a benchmark point (\textbf{BP5}) resulting in a RH sneutrino DM through coannihilation mechanism with a colored sparticle, a stop. Therefore, collider signatures on long-lived colored particles (R-hadrons) are relevant due to phase space and lack of direct sneutrino-stop coupling. Notice that in \textbf{BP4} a Bino neutralino is lighter than the stop, so in that case any stop produced in a collider would be able to decay within the detector.

In Table~\ref{BPresonances} we show benchmark points with a RH sneutrino DM using different annihilating channels. In \textbf{BP6} RH sneutrino annihilates directly to a pair of light pseudo-scalar Higgs with singlet dominant composition, via quartic coupling. This channel is especially important for direct detection experiments, and for low mass sneutrinos that can not annihilate using a resonance nor a coannihilation mechanism with Bino neutralinos.

\begin{table}
\begin{footnotesize}
\begin{center}
\begin{tabular}{|c|c|c|c|}

        \hline
        \textbf{Parameters} & \textbf{BP6} & \textbf{BP7} & \textbf{BP8}  \\
        \hline
        \hline
            $M_1$ & 1873.8 GeV & 422.9 GeV & 2418.5 GeV  \\
            $M_2$ & 782.1 GeV & 213.9 GeV & 585.9 GeV  \\
            $\mu_{eff}$ & 174.6 GeV & 137.2 GeV & 265.6 GeV  \\
            $\tan \beta$ & 13.83 & 6.152 & 4.31  \\
            $\lambda$ & 0.0750 & 0.0727 & 0.566 \\
            $\kappa$ & 0.0922 & 0.388 & 0.568 \\
            $\lambda_N^3$ & -0.114 & -0.193 & -0.413 \\
            $T_{\lambda}$ & 194.0 GeV & 0.0172 GeV & 371.6 GeV \\
            $T_{\kappa}$ & -0.0195 GeV & -0.916 GeV & -0.647 GeV  \\
            $T_{\lambda_N}^3$ & -35.33 GeV & -0.294 GeV & -12.47 GeV  \\
            $m^2_{\tilde{N}_3}$ & 4.13$\times 10^{3}$ GeV$^2$ & 9.99$\times 10^{3}$ GeV$^2$ & 7.24$\times 10^{3}$ GeV$^2$  \\
            $m^2_{\tilde{u}_3}$ & 1.80$\times 10^{6}$ GeV$^2$ & 2.67$\times 10^{6}$ GeV$^2$ & 3.08$\times 10^{6}$ GeV$^2$  \\
            $m^2_{\tilde{Q}_3}$ & 1.16$\times 10^{6}$ GeV$^2$ & 8.20$\times 10^{5}$ GeV$^2$ & 3.42$\times 10^{6}$ GeV$^2$  \\
            $T_{u_3}$ & 1854.8 GeV & 2149.4 GeV & 2250.1 GeV  \\
        \hline
        \hline
        \textbf{Spectrum} & & & \\
        \hline
        \hline
            $m_{\tilde{\nu}_{R}}$ & 89.8 GeV & 61.0 GeV & 222.1 GeV  \\
            $m_{\chi_1^0}$ & 172.1 GeV & 104.4 GeV & 247.9 GeV \\
            $m_{\chi_2^0}$ & 181.0 GeV & 146.6 GeV & 280.3 GeV \\
            $m_{\chi_1^{\pm}}$ & 175.6 GeV & 111.2 GeV & 259.9 GeV  \\
            $m_{h_1}$ & 123.2 GeV & 122.9 GeV & 125.0 GeV \\
            $m_{h_2}$ & 487.5 GeV & 786.8 GeV & 519.5 GeV  \\
            $S_{23}^2$ & 0.99 & 1.1$\times 10^{-4}$ & 0.98  \\
            $m_{A_1}$ & 21.2 GeV & 70.9 GeV & 154.9 GeV  \\
            $P_{12}^2$ & 0.99 & 0.99 & 0.99 \\
            $m_{A_2}$ & 2591.6 GeV & 786.7 GeV & 1046.7 GeV  \\
        \hline
        \hline
        \textbf{Dark Matter} & $\tilde{\nu}_{R}$ & $\tilde{\nu}_{R}$ & $\tilde{\nu}_{R}$  \\
        \hline
        \hline
            $\Omega_{\text{DM}}h^2$ & 0.107 & 6.4$\times 10^{-4}$ & 0.0167 \\
            $\sigma_{DM-p}^{SI}$ & 2.6$\times 10^{-48}$ cm$^2$ & 2.4$\times 10^{-47}$ cm$^2$ & 7.3$\times 10^{-46}$ cm$^2$ \\
            main & $\tilde{\nu}_{R} \, \tilde{\nu}_{R} \rightarrow A_1 \, A_1$ & $\tilde{\nu}_{R} \, \tilde{\nu}_{R} \rightarrow b \, \bar{b} \, (\sim 68\%)$ & $\tilde{\nu}_{R} \, \tilde{\nu}_{R} \rightarrow A_1 \, A_1 \, (\sim 65\%)$ \\
            annihilation &  & $\tilde{\nu}_{R} \, \tilde{\nu}_{R} \rightarrow g \, g \, (\sim 15\%)$ & $\tilde{\nu}_{R} \, \tilde{\nu}_{R} \rightarrow$ SM $(\sim 30\%)$   \\
            channels &  & $\tilde{\nu}_{R} \, \tilde{\nu}_{R} \rightarrow \tau \, \bar{\tau} \, (\sim 12\%)$ &  $\chi_{1}^0 \, \chi_{1}^0 \rightarrow$ SM $(\sim 4\%)$  \\
             &  & $\tilde{\nu}_{R} \, \tilde{\nu}_{R} \rightarrow c \, \bar{c} \, (\sim 4\%)$  &   \\
        \hline
\end{tabular}
\end{center}
\end{footnotesize}
  \caption{Same as \ref{BPcoannihilation1} but benchmark points with RH sneutrino DM through direct annihilation to a pair of pseudo-scalars, resonances and coanihhilations.}
  \label{BPresonances}
\end{table}

In \textbf{BP7} we present a parameter point with a $m_{\tilde{\nu}_R}\sim m_{h_1}/2$ fulfilling the resonant condition with the SM-like Higgs boson. As we can see, very low relic abundances can be achieved this way.

In \textbf{BP8} we show an example with a RH sneutrino DM that presents the three main mechanisms discussed in this work: annihilation to a pair of CP-odd Higgs, resonance with a CP-even Higgs (in this case with the second lightest Higgs that has a singlet dominant composition) and coannihilation with Higgsino dominated neutralino. Notice that solutions with different annihilation channel weights can be obtained modifying the relevant parameters, the RH sneutrino mass (e.g. $\lambda_N$), the Higgs sector mass (e.g. $\lambda$, $\kappa$), or the Higgsino mass (e.g. $\mu_{eff}$).


\pagebreak

\bibliographystyle{utphys}
\bibliography{munussm}

\end{document}

%% file: defsV3.tex
\usepackage[T1]{fontenc}
\usepackage{epsfig}
\usepackage{latexsym}
\usepackage{graphicx}
\usepackage{amsmath}
\usepackage{amsfonts}   
\usepackage{amssymb}    
\usepackage{float}
\usepackage{bm}
\usepackage{url}
\usepackage{hyperref} 
\usepackage[nodisplayskipstretch]{setspace}
\def\lsim{\raise0.3ex\hbox{$\;<$\kern-0.75em\raise-1.1ex\hbox{$\sim\;$}}}
\def\gsim{\raise0.3ex\hbox{$\;>$\kern-0.75em\raise-1.1ex\hbox{$\sim\;$}}}

\newcommand{\captions}{\sf\caption}
\def    \beq            {\begin{equation}}
\def    \eeq            {\end{equation}}
\def    \bea           {\begin{eqnarray}}
\def    \eea           {\end{eqnarray}}

\def\g2{{\rm GeV}^2}

\def\sw2{sin^2 \theta_w}

\def\a^tau{\alpha_{\tau}}

\def\beq{\begin{equation}}
\def\eeq{\end{equation}}
\def\beqa{\begin{eqnarray}}
\def\eeqa{\end{eqnarray}}

\newcommand{\newc}{\newcommand}
\newc\BR{BR}
\newc{\akappa}{A_{\kappa} }
\newc\deltagmtwo{\delta (g-2)_{\mu}} 
\newc\deltaamu{\Delta a_{\mu}}

\def\anti{\overline}

\newc{\haa}{BR\(h_1\to a_1 a_1\)}
\newc{\abb}{BR\(a_1\to b\anti{b}\)}
\newc{\hbb}{BR\(h_1\to b\anti{b}\)}
\newc{\abund}{\Omega h^2}
\newc\bsgamma{b\rightarrow s \gamma }
\newc\bxsgamma{\overline{B}\rightarrow X_{s}\gamma}
\newc\brbsgamma{\BR(\overline{B}\rightarrow X_s\gamma)}

\newc{\Fermi}{\textit{Fermi}-}
